\documentclass[aps,preprint,onecolumn,secnumarabic,nobalancelastpage,amsmath,amssymb,
nofootinbib]{revtex4}

\usepackage[hang,nooneline,scriptsize]{subfigure}
\usepackage{graphics}      
\usepackage{graphicx}      
\usepackage{longtable}     
\usepackage{pst-plot}
\usepackage{dcolumn}
\usepackage{hyperref}
\usepackage{bm}            
\usepackage{mathrsfs}

\begin{document}

\preprint{APS/123-QED}

\title{Analytic treatment of complete geodesics in a static cylindrically symmetric conformal spacetime}

\author{Bahareh Hoseini${}^1$}
\author{Reza Saffari${}^1$}%
\email{rsk@guilan.ac.ir}
\author{Saheb Soroushfar${}^1$}
\author{Saskia Grunau${}^2$}
\author{Jutta Kunz${}^2$}

\affiliation{${}^1$Department of Physics, University of Guilan, 41335-1914, Rasht, Iran.\\
${}^2$Institut f\"ur Physik, Universit\"at Oldenburg, Postfach 2503 D-26111 Oldenburg, Germany.
}

\date{\today}

\begin{abstract}
We consider the motion of test particles and light rays 
in a static cylindrically symmetric 
conformal spacetime given by Said {\it et al.} \cite{Said:2012xt}. 
We derive the equations of motion and present their analytical solutions 
in terms of the Weierstrass $\wp$ function and the Kleinian $\sigma$ function. 
Using parametric diagrams and effective potentials 
we analyze the possible orbits and characterize them in terms of the energy 
and the angular momentum of the test particles. 
Finally we show some examples of orbits.
\end{abstract}

\maketitle

\section{INTRODUCTION}

Conformal Gravity (CG)  
(see e.g.~\cite{Mannheim:2011ds})
represents an interesting alternative gravity theory
to Einstein's General Relativity (GR).
While GR has formidably passed all experimental and observational tests so far,
our understanding of the composition of galaxies and of the evolution of the Universe
within GR is based on the assumption of the existence of dark matter and dark energy,
making up 95\% of the content of the Universe.

Like GR, CG is a completely covariant metric theory of gravity.
However, CG is based on an additional symmetry principle, namely local conformal invariance. 
The presence of conformal symmetry inhibits both the Einstein-Hilbert action and a
cosmological term in the action. Instead, the action is defined in terms of
the Weyl tensor,
\begin{equation}
\label{action}
S_{\rm CG}=-\alpha_g \int d^4x \sqrt{-g}
C_{\kappa\lambda\mu\nu}C^{\kappa\lambda\mu\nu}.
\end{equation}
Being conformally invariant, the theory is sensitive to angles, but not to distances,
where the conformal transformation of the metric is given by
$g_{\mu\nu}\rightarrow \Omega^2(x)g_{\mu\nu}$.

In contrast to GR,
the gravitational coupling constant $\alpha_g$ of CG is a dimensionless constant,
making the theory power counting renormalisable. This allows to consider
CG as a quantum theory of gravity \cite{Mannheim:2011ds}.
However, 
the CG action (\ref{action}) leads to fourth order equations of motion,
which implies the presence of ghosts. 
Ways to eliminate these ghosts have been considered in 
\cite{Bender:2007wu,Maldacena:2011mk}.
On the other hand, fourth order equations of motion
imply more integration constants and thus solutions with more parameters.

A static sperically symmetric vacuum solution of CG with metric
\begin{equation}\label{metric0}
ds^2=-B(r)dt^2+\frac{dr^2}{B(r)} +r^2 ( d\theta^2 +\sin^2 \theta d \varphi^2) ,
\end{equation}
where
\begin{equation}\label{metric00}
B(r)=1 - \frac{\beta(2 - 3 \beta \gamma)}{r} - 3\beta\gamma+\gamma r-kr^2 ,
\end{equation}
was studied by Mannheim and Kazanas \cite{Mannheim:1988dj,Mannheim:1990ya}.   
Here $\beta$, $\gamma$ and $k$ are three integration constants, where the choice 
$\gamma=k=0$ yields the Schwarzschild solution, and $\gamma=0$
the Schwarzschild-de Sitter solution. Thus, $\beta$ corresponds to the
mass of the solution, while $\gamma$ characterizes the deviation
from GR. So for sufficiently small $\gamma$ (and $r$) both theories yield similar results.

On the one hand, this suggests that the Newtonian limit is reproduced, and the well-known
physics in the solar system is recovered. On the other hand, the linear term will 
not be negligable at large distance scales, allowing to fit rotation curves of galaxies
without the need for dark matter, when the parameter $\gamma$
is associated with the inverse Hubble length \cite{Mannheim:2012qw,Mannheim:2005bfa}.
At the same time, the constant $k$ acts like a cosmological constant, which, however,
enters in CG only at the level of the solutions, whereas in GR it enters as part of the action.

The solutions of the geodesic equations in a given spacetime
provide crucial information on the spacetime. For instance, 
one can infer information about the properties of a black hole from the observation
of the black hole shadow \cite{Hioki:2009na}, 
one can model the inspirals of stellar mass objects towards supermassive black holes
leading to gravitational waves models for Extreme Mass Ratio Inspirals (EMRI) 
to be observed at LISA \cite{Barack:2006pq}, 
or model the inspiralling motion of two stellar mass black holes 
employing the effective one-body
formalism  \cite{Damour:1999cr}, to find the corresponding gravitational waves models
as recently observed at LIGO  \cite{Abbott:2016blz}.
Analytic solutions of the geodesic equations allow to precisely identify
homoclinic orbits \cite{PerezGiz:2008yq}, or to
test numerical codes for binary systems, etc.
Moreover, analytic solutions can also be of use for practical applications 
like geodesy.

In 1931 Hagihara \cite{Y. Hagihara} solved the geodesic equations  
in a Schwarzschild gravitational field, where he applied the elliptic Weierstrass function.
The solutions for the Kerr and Kerr-Newman spacetimes have the same mathematical structure 
\cite{S. Chandrasekhar} and can be solved analogously. 
The mathematical method to solve the hyperelliptic equations of motion 
in the Schwarzschild-(anti) de Sitter spacetime is based on the solution of the Jacobi inversion problem restricted to the $\theta$-divisor \cite{Hackmann:2008zza,Hackmann:2008zz}. 
Also these more advanced methods were applied to obtain 
solutions of the geodesic equations 
in various spacetimes (see e.g.~\cite{Enolski:2010if,Kagramanova:2012hw,Hackmann:2009rp,Hackmann:2010ir,Grunau:2012ai,Grunau:2012ri,Grunau:2013oca}). 
Moreover, the geodesic equations were solved analytically in the spacetimes of $f(R)$ gravity, BTZ and GMGHS black holes \cite{Soroushfar:2015wqa,Soroushfar:2015dfz,Soroushfar:2016yea}. 

Here, we are interested in the geodesic equations of CG spacetimes.
For the description of the motion of stars and gas in galaxies, timelike
geodesics should be considered, in principle, though basically Newtonian
dynamics has been applied in this case
\cite{Mannheim:2012qw,Mannheim:2005bfa}.
Timelike geodesics in the static spherically symmetric CG metric 
(\ref{metric0}) have been calculated to determine the perihelion shift
of the planets in the solar system \cite{Sultana:2012qp},
where the effect of the linear term in the metric has suggested
a constraint for the integration constant $\gamma$.
Another recent calculation of timelike geodesics 
\cite{Varieschi:2014ata} has employed
the rotating generalization of the CG metric (\ref{metric0})
\cite{Mannheim:1990ya}. Exploiting separability
it has addressed the Flyby Anomaly in this CG spacetime.

We note that applicability and interpretation
of the geodesic equations and their solutions has remained a matter
of debate for CG, including considerations, that only null geodesics are
physically meaningful in CG, since they do not involve a mass scale,
while various amendments have been suggested for the desciption
of massive particles
(see e.g.~\cite{Barabash:1999bj,Edery:2001at,Wood:2001ve,Brihaye:2009ef,Ohanian:2015wva}).

In this paper, we discuss the geodesic motion of test particles 
and light in a conformal 
cylindrically symmetric spacetime obtained in \cite{Said:2012xt}.
It represents a CG generalization of the AdS black string metric obtained 
in GR by Lemos \cite{Lemos:1994xp} (for earlier work see
\cite{Linet:1986sr,Santos:1993}).
We here present the results in terms of Weierstrass elliptic functions 
and derivatives of Kleinian sigma functions. 

Our paper is organized as follows:
First, in Sec.~\ref{cg} we give a brief review of the field equations in CG,
and review some general properties of the cylindrical spacetime.
In Sec.~\ref{ge} we present the geodesic equations for this spacetime. 
In Sec.~\ref{as}, we derive the analytical solution of the equations of 
motions and describe test particle motion in this spacetime. 
We exhibit a set of possible orbits in Sec.~\ref{o} and conclude in Sec.~\ref{c}.

\section{CYLINDRICAL SOLUTION IN CONFORMAL WEYL GRAVITY}\label{cg}

The main element of CG is the substitution of the Einstein-Hilbert action 
with the Weyl action (\ref{action})
based on the Weyl tensor $C_{\kappa\lambda\mu\nu}$,
\begin{equation}\label{C}
C_{\kappa\lambda\mu\nu}=R_{\kappa\lambda\mu\nu}-\frac{1}{2}(g_{\kappa\mu}R_{\lambda\nu}-g_{\kappa\nu}R_{\lambda\mu}+g_{\lambda\nu}R_{\kappa\mu}-g_{\lambda\mu}R_{\kappa\nu})+\frac{R}{6}(g_{\kappa\mu}g_{\lambda\nu}-g_{\kappa\nu}g_{\lambda\mu}),
\end{equation}
defined as the totally traceless part of the Riemann tensor.

The CG field equations are similar to the Einstein equations, 
where the source term on the right-hand side 
is given by the energy-momentum tensor $T_{\mu\nu}$,
while on the left-hand side the Bach tensor $W_{\mu\nu}$ 
\begin{equation}
W_{\mu\nu}=\frac{1}{3}\nabla_\mu \nabla_\nu R-\nabla_\lambda \nabla^\lambda R_{\mu\nu}+\frac{1}{6}(R^2+\nabla_\lambda \nabla^\lambda R-3R_{\kappa\lambda}R^{\kappa\lambda})g_{\mu\nu} +2R^{\kappa\lambda}R_{\mu\kappa\nu\lambda}-\frac{2}{3}RR_{\mu\nu}
\end{equation}
replaces the Einstein tensor, leading to
\begin{equation}\label{bach}
2 \alpha_g W_{\mu\nu}=\frac{1}{2}T_{\mu\nu}.
\end{equation}
In vacuum, the right-hand side vanishes.

Static and stationary CG solutions were investigated 
in detail in \cite{Mannheim:1988dj,Mannheim:1990ya}. 
The case of static cylindrically symmetric solutions
was studied by Brihaye and Verbin
\cite{Brihaye:2009xc,Verbin:2010tq} and subsequently
by Said {\it et al.}~\cite{Said:2012xt}. 

Here we consider the static cylindrically symmetric
vacuum metric as given in \cite{Said:2012xt}
\begin{equation}\label{metric}
ds^2=-B(r)dt^2+\frac{dr^2}{B(r)} +r^2d\varphi^2 +\alpha^2 r^2dz^2,
\end{equation}
where $B(r)$ was derived 
by solving Eq.~(\ref{bach}) in vacuum, yielding
\begin{equation}\label{metrics}
B(r)=\frac{\beta}{r}+\sqrt{\frac{3\beta\gamma}{4}}+\frac{\gamma r}{4}+k^2r^2,
\end{equation}
with $\beta$, $\gamma$, and $k$ being integration constants.

For comparison with GR, we briefly recall the AdS black string solution
of Lemos \cite{Lemos:1994xp}, which has the metric function $B(r)$,
\begin{equation}\label{lemosm}
B(r) = \alpha^2 r^2 - \frac{b}{\alpha r}.
\end{equation}
Here $\alpha^2= - \Lambda/3 > 0$ is related to the negative cosmological
constant $\Lambda$, and $b$ is proportional to the mass, $b=M/2$.
Clearly, setting $\gamma=0$ in the CG expression (\ref{metrics}), 
we recover the GR result (\ref{lemosm}) for
\begin{equation}\label{equiv}
k = \alpha \, \ \ \ \beta=-\frac{b}{\alpha}.
\end{equation}


The metric (\ref{metrics}) possesses horizons 
when $B(r_h)=0$ \cite{Said:2012xt}.
In contrast to GR, the sign of $\Lambda$ is not predetermined here.
So for positive $\Lambda$ also a cosmological horizon may be present.
As discussed in Ref.~\cite{Verbin:2010tq}, also regular spacetimes are 
among the possible set of solutions.
Unfortunately, however, the gauge chosen in Ref.~\cite{Verbin:2010tq}
makes a direct comparison of the solutions unfeasible.
This also holds for the solutions of the null geodesics
presented in Ref.~\cite{Verbin:2010tq}.

\section{THE GEODESIC EQUATION}\label{ge}

In this section, we derive the equations of motion for test particles and light. The geodesic motion in such a spacetime of Eq.~(\ref{metric}) is described by 
\begin{equation}\label{geodesy}
\frac{d^2x^\mu}{ds^2}+\Gamma^\mu_{\rho\sigma}\frac{dx^\rho}{ds}\frac{dx^\sigma}{ds} =0,
\end{equation}
where
$\Gamma^\mu_{\rho\sigma}$ are the Christoffel symbols.
The first constant of motion is given by the normalization condition
$ds^2=\frac{1}{2}g_{\mu\nu}\frac{dx^\mu}{ds}\frac{dx^\nu}{ds}=-\frac{1}{2}\epsilon$, where
for massive particles $\epsilon=1$ and for light $\epsilon=0$.
The conserved energy and the angular momentum are
\begin{equation}\label{E}
E=-g_{tt}\frac{dt}{ds}=\frac{dt}{ds}(\frac{\beta}{r}+\sqrt{\frac{3\beta\gamma}{4}}+\frac{\gamma r}{4}+k^2r^2),
\end{equation}
\begin{equation}\label{L}
L=g_{\varphi\varphi}\frac{d\varphi}{ds}=r^2\frac{d\varphi}{ds} \, .
\end{equation}
A further constant of motion is the momentum in the $z$-direction
\begin{equation}
 J=g_{zz}\frac{dz}{ds}=\alpha^2r^2\frac{dz}{ds} \, .
\end{equation}
From Eq.~(\ref{geodesy}), we obtain
equations for $r$ as a functions of $\tau,\phi,t$ and $z$ which describe the dynamics of test particles and light
\begin{equation}\label{rtu}
(\frac{dr}{d\tau})^2=E^2-(\frac{\beta}{r}+\sqrt{\frac{3\beta\gamma}{4}}+\frac{\gamma r}{4}+k^2r^2)(\epsilon +\frac{L^2}{r^2}+\frac{J^2}{\alpha^2 r^2}),
\end{equation}
\begin{equation}\label{rphi}
(\frac{dr}{d\phi})^2=\frac{r^4}{L^2}(E^2-(\frac{\beta}{r}+\sqrt{\frac{3\beta\gamma}{4}}+\frac{\gamma r}{4}+k^2r^2)(\epsilon +\frac{L^2}{r^2}+\frac{J^2}{\alpha^2 r^2}))=R(r),
\end{equation}
\begin{equation}\label{rt}
(\frac{dr}{dt})^2=\frac{1}{E^2}(\frac{\beta}{r}+\sqrt{\frac{3\beta\gamma}{4}}+\frac{\gamma r}{4}+k^2r^2)(E^2-(\frac{\beta}{r}+\sqrt{\frac{3\beta\gamma}{4}}+\frac{\gamma r}{4}+k^2r^2)(\epsilon +\frac{L^2}{r^2}+\frac{J^2}{\alpha^2 r^2})),
\end{equation}
\begin{equation}\label{rz}
(\frac{dr}{dz})^2=\frac{\alpha^4 r^4}{J^2}(E^2-(\frac{\beta}{r}+\sqrt{\frac{3\beta\gamma}{4}}+\frac{\gamma r}{4}+k^2r^2)(\epsilon +\frac{L^2}{r^2}+\frac{J^2}{\alpha^2 r^2})).
\end{equation}
Eq.~(\ref{rtu}), suggests the introduction of an effective potential
\begin{equation}\label{veff}
V_{eff}=(\frac{\beta}{r}+\sqrt{\frac{3\beta\gamma}{4}}+\frac{\gamma r}{4}+k^2r^2)(\epsilon +\frac{L^2}{r^2}+\frac{J^2}{\alpha^2 r^2}).
\end{equation}

\section{Analytical solution of geodesic equations}\label{as}

In this section, we present the analytical solution of geodesic equations of test particles and light rays in conformal Lemos-like spacetime. We solve the  $\tilde{r}$-$\phi$-equation \eqref{rphi} and the  $\tilde{r}$-$z$-equation \eqref{rz}, which can then be used to plot the orbits. For null geodesics the solutions are given in terms of the elliptic Weierstrass $\wp$-function. The case of timelike geodesics is more complicated, here the equations are of hyperelliptic type and the Kleinian $\sigma$ function is needed to solve the equations.

\subsection{The $\tilde{r}$-$\phi$-equation}\label{req}
With the dimensionless quantities, $\tilde{r}=r/M$, $\tilde{\beta}=\beta/M$, $\tilde{\gamma}=M\gamma$, $\tilde{k}=kM, \tilde{\alpha}=M\alpha$
and $\mathcal{L}=M^2/L^2$, Eq.~(\ref{rphi}) can be written as
\begin{eqnarray}\label{Rtild}
(\frac{d\tilde{r}}{d\phi})^2 &=&-\tilde{k}^2\epsilon \mathcal{L}\tilde{r}^6-\frac{\tilde{\gamma}\epsilon\mathcal{L}}{4}\tilde{r}^5-(\sqrt{\frac{3\tilde{\beta}\tilde{\gamma}}{4}}\epsilon \mathcal{L}+k^2+\frac{\tilde{k}^2 J^2\mathcal{L}}{\tilde{\alpha}^2}-E^2\mathcal{L})\tilde{r}^4 \nonumber \\
&-&(\epsilon \tilde{\beta} \mathcal{L}+\frac{\tilde{\gamma}}{4}+\frac{\tilde{\gamma} J^2 \mathcal{L}}{4\tilde{\alpha}^2})\tilde{r}^3  -(\sqrt{\frac{3\tilde{\beta}\tilde{\gamma}}{4}}+\sqrt{\frac{3\tilde{\beta}\tilde{\gamma}}{4}}\frac{J^2\mathcal{L}}{\tilde{\alpha}^2})\tilde{r}^2\nonumber  \\
&-&(\tilde{\beta}+\frac{\tilde{\beta} J^2 \mathcal{L}}{\tilde{\alpha}^2})\tilde{r}=R(\tilde{r}).
\end{eqnarray}
Eq.~(\ref{Rtild}) implies that $R(\tilde{r})\geq 0$ is a necessary condition for the existence of a geodesic. We also observe that $\tilde{r} = 0$, where the singularity is located, is a zero of $R(\tilde{r})$ for all values of the parameters. The real and positive zeros of $R(\tilde{r})$ are the turning points of the geodesics and determine the possible types of orbits.

In general $R(\tilde{r})$ is a polynomial of order six, but in the special case $\epsilon=0$ it simplifies to order four. Therefore we will treat null geodesics and timelike geodesics separately.

\subsubsection{Null geodesics}
For $\epsilon=0$, Eq.~(\ref{Rtild}) is of elliptic type. The polynomial  $R(\tilde{r})$ can be reduced to third order by substituting $\tilde{r}=\frac{1}{u}$
\begin{eqnarray}\label{p3}
(\frac{du}{d\varphi})^2&=&-(\tilde{\beta}+\frac{\tilde{\beta} {J}^2 \mathcal{L}}{\tilde{\alpha}^2})u^3-(\sqrt{\frac{3\tilde{\beta}\tilde{\gamma}}{4}}+\sqrt{\frac{3\tilde{\beta}\tilde{\gamma}}{4}}\frac{{J}^2\mathcal{L}}{\tilde{\alpha}^2})u^2-(\epsilon \tilde{\beta} \mathcal{L}+\frac{\tilde{\gamma}}{4}+\frac{\tilde{\gamma} {J}^2 \mathcal{L}}{4\tilde{\alpha}^2})u\nonumber  \\
&-&(\tilde{k}^2+\frac{\tilde{k}^2 {J}^2\mathcal{L}}{\tilde{\alpha}^2}-E^2\mathcal{L})=P_3(u)=\sum_{i=0}^3 a_i u^i.
\end{eqnarray}
A further substitution
\begin{equation}
u=\frac{1}{a^3}(4y-\frac{a_{2}}{3})=\frac{1}{(\tilde{\beta}+\frac{\tilde{\beta} {J}^2 \mathcal{L}}{\tilde{\alpha}^2})}(4y+\frac{1}{3}(\sqrt{\frac{3\tilde{\beta}\tilde{\gamma}}{4}}+\sqrt{\frac{3\tilde{\beta}\tilde{\gamma}}{4}}\frac{{J}^2\mathcal{L}}{\tilde{\alpha}^2})),
\end{equation}
transforms $P_3(u)$ into the Weierstrass form, so that Eq.~(\ref{p3}) turns into:
\begin{equation}\label{p31y}
(\frac{dy}{d\varphi})^2=4y^3-g_2y-g_3=P_3(y),
\end{equation}
with
\begin{equation}
g_2=\frac{a_2^2}{12}-\frac{a_1a_3}{4}=\frac{1}{12}(\sqrt{\frac{3\tilde{\beta}\tilde{\gamma}}{4}}+\sqrt{\frac{3\tilde{\beta}\tilde{\gamma}}{4}}\frac{{J}^2\mathcal{L}}{\tilde{\alpha}^2})^2-\frac{1}{4}(\epsilon \tilde{\beta} \mathcal{L}+\frac{\tilde{\gamma}}{4}+\frac{\tilde{\gamma} {J}^2 \mathcal{L}}{4\tilde{\alpha}^2})(\tilde{\beta}+\frac{\tilde{\beta} {J}^2 \mathcal{L}}{\tilde{\alpha}^2}),
\end{equation}
\begin{eqnarray}
g_3&=&\frac{a_1a_2a_3}{48}-\frac{a_0a_3^2}{16}-\frac{a_2^3}{216}=-\frac{1}{216}(\sqrt{\frac{3\tilde{\beta}\tilde{\gamma}}{4}}+\sqrt{\frac{3\tilde{\beta}\tilde{\gamma}}{4}}\frac{J^2\mathcal{L}}{\tilde{\alpha}^2})^2\nonumber  \\
&-&(\epsilon \tilde{\beta} \mathcal{L}+\frac{\tilde{\gamma}}{4}+\frac{\tilde{\gamma} {J}^2 \mathcal{L}}{4\tilde{\alpha}^2})(\sqrt{\frac{3\tilde{\beta}\tilde{\gamma}}{4}}+\sqrt{\frac{3\tilde{\beta}\tilde{\gamma}}{4}}\frac{{J}^2\mathcal{L}}{\tilde{\alpha}^2})(\tilde{\beta}+\frac{\tilde{\beta} {J}^2 \mathcal{L}}{\tilde{\alpha}^2})\nonumber  \\
&+&\frac{1}{16}(\tilde{k}^2+\frac{\tilde{k}^2 {J}^2\mathcal{L}}{\tilde{\alpha}^2}-E^2\mathcal{L})(\tilde{\beta}+\frac{\tilde{\beta} {J}^2 \mathcal{L}}{\tilde{\alpha}^2})^2.
\end{eqnarray}
Eq.~(\ref{p31y}) is solved by the Weierstrass function \cite{Hackmann:2008zz,M.Abramowitz,E. T. Whittaker},
\begin{equation}\label{wp}
y(\varphi)=\wp(\varphi-\varphi_{in};g_2,g_3),
\end{equation}
where, $\varphi_{in}=\varphi_0+\int_{y_0}^\infty \frac{dy}{\sqrt{4y^3-g_2y-g_3}}$,
with
\begin{equation}
y_0=\frac{a_3}{4\tilde{r}_0}+\frac{a_2}{12}=-\frac{1}{4\tilde{r}_0}(\tilde{\beta}+\frac{\tilde{\beta} {J}^2 \mathcal{L}}{\tilde{\alpha}^2})-\frac{1}{12}(\sqrt{\frac{3\tilde{\beta}\tilde{\gamma}}{4}}+\sqrt{\frac{3\tilde{\beta}\tilde{\gamma}}{4}}\frac{J^2\mathcal{L}}{\tilde{\alpha}^2}).
\end{equation}
Then the solution of Eq.~(\ref{Rtild}) in the case $\epsilon=0$ acquires the form
\begin{equation}
\tilde{r}(\varphi)=\frac{a_3}{4\wp(\varphi-\varphi_{in};g_2,g_3)-\frac{a_2}{3}}=\frac{-(\tilde{\beta}+\frac{\tilde{\beta} {J}^2 \mathcal{L}}{\tilde{\alpha}^2})}{2\wp(\varphi-\varphi_{in};g_2,g_3)+\frac{1}{3}(\sqrt{\frac{3\tilde{\beta}\tilde{\gamma}}{4}}+\sqrt{\frac{3\tilde{\beta}\tilde{\gamma}}{4}}\frac{{J}^2\mathcal{L}}{\tilde{\alpha}^2})}.
\end{equation}

\subsubsection{Timelike geodesics}\label{ssTg}
Considering the case $\epsilon=1$, Eq.~(\ref{Rtild}) is of hyperelliptic type. Using the substitution $\tilde{r}=\frac{1}{u}$ it be rewritten as
\begin{eqnarray}\label{p5}
(u\frac{du}{d\varphi})^2&=&-(\tilde{\beta}+\frac{\tilde{\beta} {J}^2 \mathcal{L}}{\tilde{\alpha}^2})u^5-(\sqrt{\frac{3\tilde{\beta}\tilde{\gamma}}{4}}+\sqrt{\frac{3\tilde{\beta}\tilde{\gamma}}{4}}\frac{{J}^2\mathcal{L}}{\tilde{\alpha}^2})u^4-(\epsilon \tilde{\beta} \mathcal{L}+\frac{\tilde{\gamma}}{4}+\frac{\tilde{\gamma} {J}^2 \mathcal{L}}{4\tilde{\alpha}^2})u^3\nonumber  \\
&-&(\sqrt{\frac{3\tilde{\beta}\tilde{\gamma}}{4}}\epsilon \mathcal{L}+\tilde{k}^2+\frac{\tilde{k}^2 {J}^2\mathcal{L}}{\tilde{\alpha}^2}-E^2\mathcal{L})u^2-\frac{\tilde{\gamma}\epsilon\mathcal{L}}{4}u-\tilde{k}^2\epsilon \mathcal{L}\nonumber  \\
&=&P_5(u)=\sum_{i=0}^5 a_i u^i.
\end{eqnarray}
This problem is a special case of the Jacobi inversion problem and can be solved when restricted to the $\theta$ divisor, the set of zeros of the Riemann $\theta$ function. The solution procedure is extensively discussed in e.g. \cite{Hackmann:2008zz,Enolski:2010if}. The analytic solution of Eq.~(\ref{p5}) is given in terms of derivatives of the Kleinian $\sigma$ function
\begin{equation}
	u(\varphi) = \left. \frac{\sigma_1 (\boldsymbol{\varphi}_\infty)}{\sigma_2 (\boldsymbol{\varphi}_\infty)} \right| _{ \sigma (\boldsymbol{\varphi}_\infty)=0} \, ,
\end{equation}
with
\begin{equation}
	\boldsymbol{\varphi}_\infty = 
	\left( 
	\begin{array}{c}
	  \varphi_2 \\
   	  \varphi-\varphi_{\rm in}'
	\end{array}
	\right),
\end{equation}
and $ \varphi_{\rm in}'=  \varphi_{\rm in}+\int_{ \varphi_{\rm in}}^{\infty}\! \frac{u \, \mathrm{d} u'}{\sqrt{ P_5(u')}}$. The component $ \varphi_2$ is determined by the condition $\sigma (\boldsymbol{\varphi}_\infty)=0$.
The function $\sigma_i$ is the $i$th derivative of Kleinian $\sigma$
function and $\sigma_z$ is
\begin{equation}
\sigma(z)=Ce^{zt}kz \theta [g,h](2\omega^{-1}z;\tau),
\end{equation}
which is given by the Riemann $\theta$-function with characteristic $[g,h]$. A number of parameters
enter here: the symmetric Riemann matrix $\tau$ , the period-matrix $(2\omega, 2\acute{\omega})$, the periodmatrix of the second kind $(2\eta, 2\acute{\eta})$, the matrix $\kappa = \eta(2ω)^{-1}$
and the vector of Riemann
constants with base point at infinity $2[g,h] = (0, 1)^t+(1,1)^t\tau$. The constant $C $, can be
given explicitly, see e.g.\cite{V. M. Buchstaber}, but does not matter here. 

Finally the analytical solution of Eq.~(\ref{Rtild}) is
\begin{equation}
r(\varphi) = \left. \frac{\sigma_2 (\boldsymbol{\varphi}_\infty)}{\sigma_1 (\boldsymbol{\varphi}_\infty)}\right| _{ \sigma (\boldsymbol{\varphi}_\infty)=0} \, .
\end{equation}
This is the analytic solution of the equation of motion
of a test particle in cylindrical  
space time in coformalgravity. The solution is valid in all regions of this spacetime.

\subsection{The $\tilde{r}$-$z$-equation}
Again, with the subsitution, $\tilde{r}=r/M$, $\tilde{\beta}=\beta/M$, $\tilde{\gamma}=M\gamma$, $\tilde{k}=kM, \tilde{\alpha}=M\alpha$
and $\mathcal{L}=M^2/L^2$, the $z$-equation (\ref{rz}) becomes
\begin{eqnarray}\label{Q}
(\frac{d\tilde{r}}{dz})^2&=&-(\frac{\tilde{\alpha}^4 \tilde{k}^2 \varepsilon}{{J}^2})\tilde{r}^6-(\frac{\tilde{\alpha}^4 \tilde{\gamma} \varepsilon}{4{J}^2})\tilde{r}^5-(\frac{\tilde{\alpha}^4 E^2}{{J}^2}-\frac{\tilde{\alpha}^4 \tilde{k}^2}{\mathcal{L}{J}^2}-\tilde{\alpha}^2 \tilde{k}^2+\frac{\tilde{\alpha}^4 \varepsilon}{\tilde{J}^2}\sqrt{\frac{3\tilde{\beta}\tilde{\gamma}}{4}})\tilde{r}^4\nonumber \\
&-&(\frac{\tilde{\alpha}^2 \tilde{\gamma}}{4\mathcal{L}{J}^2}+\frac{\tilde{\alpha}^2 \tilde{\gamma}}{4}+\frac{\tilde{\alpha}^4\tilde{\beta} \varepsilon}{{J}^2})\tilde{r}^3 -(\frac{\tilde{\alpha}^4}{{J}^2\mathcal{L}}\sqrt{\frac{3\tilde{\beta}\tilde{\gamma}}{4}}+\tilde{\alpha}^2 \sqrt{\frac{3\tilde{\beta}\tilde{\gamma}}{4}})\tilde{r}^2\nonumber \\
&-&(\frac{\tilde{\alpha}^2\tilde{\beta}}{\mathcal{L}{J}^2}+\tilde{\alpha}^2 \tilde{\beta})\tilde{r}=Q(\tilde{r})
\end{eqnarray}
$Q(\tilde{r})$ is a polynomial of order six if $\epsilon=1$ and of order four if $\epsilon=0$. For all values of the parameters $\tilde{r}=0$ as a zero of $Q(\tilde{r})$. As before we will treat null and timelike geodesics separately. The solutions can be found analogously to section \ref{req}.

\subsubsection{Null geodesics}
For $\varepsilon=0$, Eq.~(\ref{Q}) is of elliptic type and the polynomial $Q(\tilde{r})$ can be reduced to third order by substituting $\xi =\frac{1}{\tilde{r}}$
\begin{eqnarray}\label{uz0}
(\frac{d\xi}{dz})^2&=&-(\frac{\tilde{\alpha}^2\tilde{\beta}}{\mathcal{L}{J}^2}+\tilde{\alpha}^2 \tilde{\beta})\xi ^3-(\frac{\tilde{\alpha}^4}{{J}^2\mathcal{L}}\sqrt{\frac{3\tilde{\beta}\tilde{\gamma}}{4}}+\tilde{\alpha}^2 \sqrt{\frac{3\tilde{\beta}\tilde{\gamma}}{4}})\xi^2-(\frac{\tilde{\alpha}^2 \tilde{\gamma}}{4\mathcal{L}{J}^2}+\frac{\tilde{\alpha}^2 \tilde{\gamma}}{4})\xi\nonumber \\
&-&(\frac{\tilde{\alpha}^4 E^2}{J^2}-\frac{\tilde{\alpha}^4 \tilde{k}^2}{\mathcal{L}{J}^2}-\tilde{\alpha}^2 \tilde{k}^2)=P_3(\xi)=\sum_{i=0}^3 a_i \xi^i.
\end{eqnarray}
A further substitution
\begin{equation}
\xi=\frac{1}{a^3}(4y-\frac{a_{2}}{3})=\frac{-1}{(\frac{\tilde{\alpha}^2\tilde{\beta}}{\mathcal{L}{J}^2}+\tilde{\alpha}^2 \tilde{\beta})}(4y+\frac{(\frac{\tilde{\alpha}^4}{{J}^2\mathcal{L}}\sqrt{\frac{3\tilde{\beta}\tilde{\gamma}}{4}}+\tilde{\alpha}^2 \sqrt{\frac{3\tilde{\beta}\tilde{\gamma}}{4}})}{3}),
\end{equation}
transforms $P_3(\xi)$ into the Weierstrass form, so that Eq.~(\ref{uz0}) turns into:
\begin{equation}\label{p32y}
(\frac{dy}{dz})^2=4y^3-g_2y-g_3=P_3(y),
\end{equation}
with
\begin{equation}
g_2=\frac{a_2^2}{12}-\frac{a_1a_3}{4}=\frac{1}{12}(\frac{\tilde{\alpha}^4}{{J}^2\mathcal{L}}\sqrt{\frac{3\tilde{\beta}\tilde{\gamma}}{4}}+\tilde{\alpha}^2 \sqrt{\frac{3\tilde{\beta}\tilde{\gamma}}{4}})^2-\frac{1}{4}(\frac{\tilde{\alpha}^2 \tilde{\gamma}}{4\mathcal{L}{J}^2}+\frac{\tilde{\alpha}^2 \tilde{\gamma}}{4}+\frac{\alpha^4\beta \varepsilon}{J^2})(\frac{\tilde{\alpha}^2\tilde{\beta}}{\mathcal{L}{J}^2}+\tilde{\alpha}^2 \tilde{\beta}),
\end{equation}

\begin{eqnarray}
g_3&=&\frac{a_1a_2a_3}{48}-\frac{a_0a_3^2}{16}-\frac{a_2^3}{216}=-\frac{1}{216}(\frac{\tilde{\alpha}^4}{{J}^2\mathcal{L}}\sqrt{\frac{3\tilde{\beta}\tilde{\gamma}}{4}}+\tilde{\alpha}^2 \sqrt{\frac{3\tilde{\beta}\tilde{\gamma}}{4}})^3\nonumber  \\
&-&\frac{1}{48}(\frac{\tilde{\alpha}^2\tilde{\beta}}{\mathcal{L}{J}^2}+\tilde{\alpha}^2 \tilde{\beta})(\frac{\tilde{\alpha}^4}{{J}^2\mathcal{L}}\sqrt{\frac{3\tilde{\beta}\tilde{\gamma}}{4}}+\tilde{\alpha}^2 \sqrt{\frac{3\tilde{\beta}\tilde{\gamma}}{4}})(\frac{\tilde{\alpha}^2 \tilde{\gamma}}{4\mathcal{L}{J}^2}+\frac{\tilde{\alpha}^2 \tilde{\gamma}}{4}+\frac{\tilde{\alpha}^4\tilde{\beta} \varepsilon}{J^2})\nonumber  \\
&+&\frac{1}{16}(\frac{\tilde{\alpha}^4 E^2}{J^2}-\frac{\tilde{\alpha}^4 \tilde{k}^2}{\mathcal{L}{J}^2}-\tilde{\alpha}^2 \tilde{k}^2)(\frac{\tilde{\alpha}^2\tilde{\beta}}{\mathcal{L}{J}^2}+\tilde{\alpha}^2 \tilde{\beta})^2.
\end{eqnarray}
Eq.~(\ref{p32y}) is solved
by the Weierstrass function \cite{Hackmann:2008zz,M.Abramowitz,E. T. Whittaker}
\begin{equation}\label{wp2}
y(z)=\wp(z-z_{in};g_2,g_3),
\end{equation}
where $z_{in}=z_0+\int_{y_0}^\infty \frac{dy}{\sqrt{4y^3-g_2y-g_3}}$
with
\begin{equation}
y_0=\frac{a_3}{4\tilde{r}_0}+\frac{a_2}{12}=-\frac{1}{4\tilde{r}_0}(\frac{\tilde{\alpha}^2\tilde{\beta}}{\mathcal{L}\tilde{J}^2}+\tilde{\alpha}^2 \tilde{\beta})-\frac{1}{12}(\frac{\tilde{\alpha}^4}{\tilde{J}^2\mathcal{L}}\sqrt{\frac{3\tilde{\beta}\tilde{\gamma}}{4}}+\tilde{\alpha}^2 \sqrt{\frac{3\tilde{\beta}\tilde{\gamma}}{4}}).
\end{equation}
Then the solution of Eq.~(\ref{Q}) in the case $\epsilon=0$ acquires the form
\begin{equation}
\tilde{r}(z)=\frac{a_3}{4\wp(z-z_{in};g_2,g_3)-\frac{a_2}{3}}=\frac{-(\frac{\tilde{\alpha}^2\tilde{\beta}}{\mathcal{L}{J}^2}+\tilde{\alpha}^2 \tilde{\beta})}{2\wp(z-z_{in};g_2,g_3)+\frac{1}{3}(\frac{\tilde{\alpha}^4}{{J}^2\mathcal{L}}\sqrt{\frac{3\tilde{\beta}\tilde{\gamma}}{4}}+\tilde{\alpha}^2 \sqrt{\frac{3\tilde{\beta}\tilde{\gamma}}{4}})}.
\end{equation}

\subsubsection{Timelike geodesics}
For $\epsilon=1$, Eq.~(\ref{Q}) is of hyperelliptic type. With the substitution $\xi =\frac{1}{\tilde{r}}$ it can be rewritten as
\begin{eqnarray}\label{ut}
(\xi\frac{d\xi}{dz})^2&=&-(\frac{\tilde{\alpha}^2\tilde{\beta}}{\mathcal{L}{J}^2}+\tilde{\alpha}^2 \tilde{\beta})\xi^5-(\frac{\tilde{\alpha}^4}{{J}^2\mathcal{L}}\sqrt{\frac{3\tilde{\beta}\tilde{\gamma}}{4}}+\tilde{\alpha}^2 \sqrt{\frac{3\tilde{\beta}\tilde{\gamma}}{4}})\xi^4-(\frac{\tilde{\alpha}^2 \tilde{\gamma}}{4\mathcal{L}{J}^2}+\frac{\tilde{\alpha}^2 \tilde{\gamma}}{4}+\frac{\alpha^4\beta \varepsilon}{J^2})\xi^3\nonumber \\
&-&(\frac{\tilde{\alpha}^4 E^2}{{J}^2}-\frac{\tilde{\alpha}^4 \tilde{k}^2}{\mathcal{L}{J}^2}-\tilde{\alpha}^2 \tilde{k}^2+\frac{\tilde{\alpha}^4 \varepsilon}{{J}^2}\sqrt{\frac{3\tilde{\beta}\tilde{\gamma}}{4}})\xi^2-(\frac{\tilde{\alpha}^4 \tilde{\gamma} \varepsilon}{4{J}^2})\xi-(\frac{\tilde{\alpha}^4 \tilde{k}^2 \varepsilon}{{J}^2})\nonumber \\
&=&P_5(\xi)=\sum_{i=0}^5 b_i \xi^i
\end{eqnarray}
Analogously to section \ref{req} we can write the analytic solution of Eq.~(\ref{ut}) as in e.g. \cite{Hackmann:2008zz,Enolski:2010if}
\begin{equation}
	\xi(z) = \left. \frac{\sigma_1 (\boldsymbol{z}_\infty)}{\sigma_2 (\boldsymbol{z}_\infty)} \right| _{ \sigma (\boldsymbol{z}_\infty)=0} \, ,
\end{equation}
with
\begin{equation}
	\boldsymbol{z}_\infty = 
	\left( 
	\begin{array}{c}
	  z_2 \\
   	  z-z_{\rm in}'
	\end{array}
	\right)
\end{equation}
and $ z_{\rm in}'=  z_{\rm in}+\int_{ z_{\rm in}}^{\infty}\! \frac{\xi \, \mathrm{d} \xi'}{\sqrt{ P_5(\xi')}}$. The component $ z_2$ is determined by the condition $\sigma (\boldsymbol{z}_\infty)=0$. 

Finally the analytical solution of Eq.~(\ref{Q}) is
\begin{equation}
r(z) = \left. \frac{\sigma_2 (\boldsymbol{z}_\infty)}{\sigma_1 (\boldsymbol{z}_\infty)}\right| _{ \sigma (\boldsymbol{z}_\infty)=0} \, .
\end{equation}

\section{ORBITS}\label{o}

In this section, we analyze the possible orbits and characterize them in terms of the parameters of the metric and the test particles. Therefore we use parametric diagrams and effective potentials. Finally we show some example plots of the possible orbits, which are \emph{escape orbits} (EO) that approach the black hole and then escape its gravity, \emph{bound orbits} (BO) that move between two turning points, and \emph{terminating orbits} that end in the singularity at $\tilde{r}=0$. Here, we distinguish between terminating escape orbits (TEO) and terminating bound orbits (TBO).
To analyze the possible orbits we consider the $\tilde{r}$-$\phi$-equation
\begin{equation}
 (\frac{d\tilde{r}}{d\phi})^2=R(\tilde{r}),
\end{equation}
with
\begin{eqnarray}
R(\tilde{r}) &=&-\tilde{k}^2\epsilon \mathcal{L}\tilde{r}^6-\frac{\tilde{\gamma}\epsilon\mathcal{L}}{4}\tilde{r}^5-(\sqrt{\frac{3\tilde{\beta}\tilde{\gamma}}{4}}\epsilon \mathcal{L}+k^2+\frac{\tilde{k}^2 J^2\mathcal{L}}{\tilde{\alpha}^2}-E^2\mathcal{L})\tilde{r}^4 \nonumber \\
&-&(\epsilon \tilde{\beta} \mathcal{L}+\frac{\tilde{\gamma}}{4}+\frac{\tilde{\gamma} J^2 \mathcal{L}}{4\tilde{\alpha}^2})\tilde{r}^3  -(\sqrt{\frac{3\tilde{\beta}\tilde{\gamma}}{4}}+\sqrt{\frac{3\tilde{\beta}\tilde{\gamma}}{4}}\frac{J^2\mathcal{L}}{\tilde{\alpha}^2})\tilde{r}^2\nonumber  \\
&-&(\tilde{\beta}+\frac{\tilde{\beta} J^2 \mathcal{L}}{\tilde{\alpha}^2})\tilde{r}.
\end{eqnarray}
The polynomial $R$, determines the possible orbit types, since its zeros are the turning points of the geodesics.
The number of zeros changes, if double zeros appear, that is
\begin{equation}\label{condish}
R(\tilde{r})=0,\qquad\qquad \frac{dR}{d\tilde{r}}=0.
\end{equation}
From  Eq.~(\ref{condish}), we obtain two conditions
\begin{eqnarray}\label{EL}
\mathcal{L}&=&-\frac{\tilde{\alpha} (\tilde{r}^2\tilde{\gamma} +4 \tilde{r}\sqrt{3\tilde{\beta}\tilde{\gamma}}+12\tilde{\beta})}{-8\tilde{\alpha}^2 \tilde{r}^5-\tilde{\gamma}\tilde{\alpha} \tilde{r}^4+4\tilde{\alpha}\tilde{\beta} \tilde{r}^2+\tilde{\gamma} J^2\tilde{r}^2+4\sqrt{3\tilde{\beta}\tilde{\alpha}}J^2\tilde{r}+12\tilde{\beta} J^2},\nonumber \\
E^2&=&16\tilde{\beta} \sqrt{3\tilde{\beta}\tilde{\gamma}}\tilde{r}+16\tilde{\alpha} \sqrt{3\tilde{\beta}\tilde{\gamma}}\tilde{r}^4+4\tilde{\gamma} \sqrt{3\tilde{\beta}\tilde{\gamma}}\tilde{r}^3+20\tilde{\beta}\tilde{\gamma} \tilde{r}^2+16\tilde{\beta}^2 +16\tilde{\alpha}^2 \tilde{r}^6 \\
 &+&32\tilde{\alpha}\tilde{\beta} \tilde{r}^3+\tilde{\gamma}^2 \tilde{r}^4+8\tilde{\alpha}\tilde{\gamma} \tilde{r}^5/2\tilde{r}(\tilde{\gamma} \tilde{r}^2+4\sqrt{3\tilde{\beta}\tilde{\gamma}}\tilde{r}+12\tilde{\beta}), \nonumber 
\end{eqnarray}
which can be used to draw parametric diagrams, that divide the $E^2$-$\mathcal{L}$-plane into several regions.

As $\tilde{r} = 0$, is a zero of $ R(\tilde{r})$ for all values of
the parameters, it is neglected in the following analysis and
\begin{eqnarray}\label{Rtilds}
R^{*}(\tilde{r}) &=&-\tilde{k}^2\epsilon \mathcal{L}\tilde{r}^5-\frac{\tilde{\gamma} \epsilon\mathcal{L}}{4}\tilde{r}^4-(\sqrt{\frac{3\tilde{\beta}\tilde{\gamma}}{4}}\epsilon \mathcal{L}+\tilde{k}^2+\frac{\tilde{k}^2 {J}^2\mathcal{L}}{\tilde{\alpha}^2}-E^2\mathcal{L})\tilde{r}^3\nonumber  \\
&-&(\epsilon \tilde{\beta} \mathcal{L}+\frac{\tilde{\gamma}}{4}+\frac{\tilde{\gamma}{J}^2 \mathcal{L}}{4\tilde{\alpha}^2})\tilde{r}^2  -(\sqrt{\frac{3\tilde{\beta}\tilde{\gamma}}{4}}+\sqrt{\frac{3\tilde{\beta}\tilde{\gamma}}{4}}\frac{J^2\mathcal{L}}{\tilde{\alpha}^2})\tilde{r}\nonumber  \\
&-&(\tilde{\beta}+\frac{\tilde{\beta} { J}^2 \mathcal{L}}{\tilde{\alpha}^2}),
\end{eqnarray}
is considered instead.

\subsection{Special case $k=\alpha$}

First, we investigate the special case $k=\alpha$ to compare with the general relativistic solution $\gamma=0$ where $\alpha=k=\sqrt{-\Lambda}$ is related to the cosmological constant.

\begin{figure}[ht]
\centerline{\includegraphics[width=7cm]{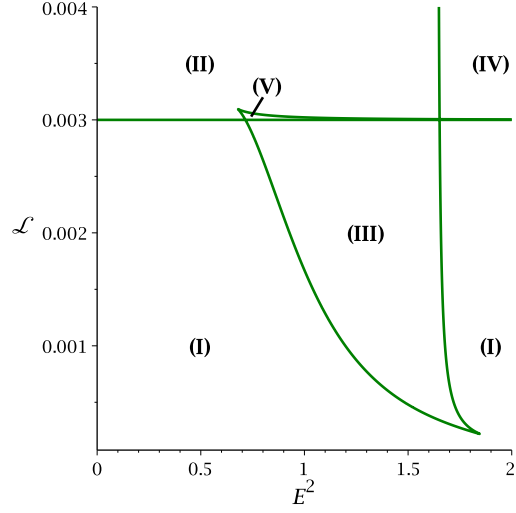}}
\caption{\label{el}\small   
 Parametric $\mathcal{L}$-$E^2$-diagram with the parameters $\varepsilon=1,\tilde{\beta}=3,\tilde{\gamma}=0.05,J=0.1, \tilde{\alpha}^2=\tilde{k}^2=-\Lambda=-3\cdot 10^{-5}$. There are five regions with a different number of zeros (see text).}
\end{figure}

\begin{figure}[h!]
\centerline{\includegraphics[width=7cm]{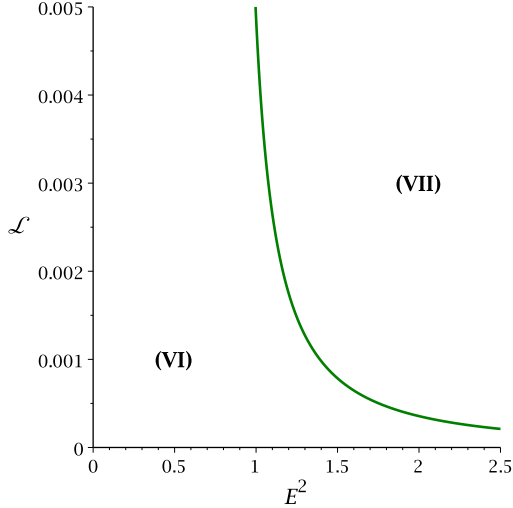}}
\caption{\label{el2}\small   
 Parametric $\mathcal{L}$-$E^2$-diagram with the parameters $\varepsilon=1,\tilde{\beta}=1,\tilde{\gamma}=0.05,J=0.1, \tilde{\alpha}^2=\tilde{k}^2=-\Lambda=3\cdot 10^{-5}$. There are two regions with a different number of zeros (see text).}
\end{figure}

In Fig.~\ref{el} and Fig.~\ref{el2} we show parametric $\mathcal{L}$-$E^2$-diagrams based on Eq.~(\ref{EL}). Up to five regions with a different number of zeros can be distinguished. In Fig.~\ref{el} the cosmological constant is positive and in Fig.~\ref{el2} it is negative.
Additionally we consider the effective potential in each region given by Eq. \eqref{veff} to visualize the orbits. Some plots of the effective potential with energies corresponding to certain orbits are depicted in Fig.~\ref{v1}.

Taking all the information into account, we find all possible orbits in the static cylindrically symmetric conformal spacetime. 

First, we consider the case of a \emph{positive cosmological constant} $\Lambda>0$. In this case, there is no event horizon, so that the singularity is naked. In the parametric $\mathcal{L}$-$E^2$-diagram (Fig.~\ref{el}) we recognize five regions with a different number of zeros. This number also depends on the sign of $\beta$ and $\gamma$. To obtain real values in the function $B(r)$ either  $\beta, \gamma >0$ or  $\beta, \gamma <0$ can be chosen. If the sign of $\beta \gamma$ is reversed, then the polynomial $R(r)$ (or $R^*(\tilde{r})$) is mirrored with respect to the ordinate so that all zeros change their sign. Since the curvature singularity is at $\tilde{r}=0$, we are interested in the positive zeros only. Table \ref{tab:cyl.orbits}, shows an overview of the different regions and the possible orbit types in following regions (below we assume that $r_i<r_{i+1}$). 
\begin{enumerate}
	\item Region I
	\begin{enumerate}
		\item $\beta, \gamma >0$: $R^*(\tilde{r})$ has a single positive zero $r_1$ and $R^*(\tilde{r})>0$ for $\tilde{r}\in [r_1,\infty)$ . Here only an escape orbit exists.
		\item  $\beta, \gamma <0$: There are no positive zeros and $R^*(\tilde{r})>0$ for all $\tilde{r}\geq0$. The corresponding orbit is a terminating escape orbit.
	\end{enumerate}

	\item Region II
	\begin{enumerate}
		\item $\beta, \gamma >0$: $R^*(\tilde{r})$ has two positive zeros $r_1$ and $r_2$. $R^*(\tilde{r})>0$ for $\tilde{r}\in [0,r_1]$ and for $\tilde{r}\in [r_2,\infty)$. Therefore, terminating bound orbits and escape orbits are possible.
		\item  $\beta, \gamma <0$: $R^*(\tilde{r})$ has a single positive zero $r_1$ and $R^*(\tilde{r})>0$ for $\tilde{r}\in [r_1,\infty)$ . Here an escape orbit exists.
	\end{enumerate}
	\item Region III
	\begin{enumerate}
		\item $\beta, \gamma >0$: $R^*(\tilde{r})$ has three positive real zeros $r_1$, $r_2$, $r_3$ and $R^*(\tilde{r})>0$ for $\tilde{r}\in [r_1,r_2]$ and for  $\tilde{r}\in [r_3,\infty)$. Here we find bound orbits and escape orbits.
		\item  $\beta, \gamma <0$:  There are no positive zeros and $R^*(\tilde{r})>0$ for all $\tilde{r}\geq0$. The corresponding orbit is a terminating escape orbit.
	\end{enumerate}

	\item Region IV
	\begin{enumerate}
		\item $\beta, \gamma >0$: There are no positive zeros and $R^*(\tilde{r})>0$ for all $\tilde{r}\geq0$. The corresponding orbit is a terminating escape orbit.
		\item  $\beta, \gamma <0$: $R^*(\tilde{r})$ has a single positive zero $r_1$ and $R^*(\tilde{r})>0$ for $\tilde{r}\in [r_1,\infty)$ . Here an escape orbit exists.
	\end{enumerate}

	\item Region V
	\begin{enumerate}
		\item $\beta, \gamma >0$: There are four positive zeros $r_1$, $r_2$, $r_3$, $r_4$ and $R^*(\tilde{r})>0$ for $\tilde{r}\in (0,r_1]$, for $\tilde{r}\in [r_2,r_3]$ and for  $\tilde{r}\in [r_4,\infty)$. The corresponding orbits are, terminating escape orbit, bound orbit and escape orbit.  
		\item  $\beta, \gamma <0$: $R^*(\tilde{r})$ has a single positive zero $r_1$ and $R^*(\tilde{r})>0$ for $\tilde{r}\in [r_1,\infty)$ . Here an escape orbit exists.
	\end{enumerate}
\end{enumerate}
Note that for lightlike geodesics $\epsilon=0$, only the regions I,II and IV are present. Furthermore, for $\epsilon=0$  the number of zeros in region II changes. For $\beta, \gamma >0$ a single zero exists in region II so that only TBOs are possible and bound orbits do not exist for lightlike geodesics. For $\beta, \gamma <0$ geodesic motion is not possible at all in region II. 

In the corresponding GR case $\gamma=0$, a solution with $\Lambda>0$ doesn't exist. Therefore $\Lambda>0$ solutions and the orbit configurations shown in Tab. \ref{tab:cyl.orbits} are features of CG. However, in the case $\alpha^2=k^2=-\Lambda <0$ a negative $\alpha^2$ negative makes the $z$ coordinate a timelike coordinate. We include this case for the sake of completeness and mathematical curiosity. 
\\

\begin{table}[h]
\begin{center}
\begin{tabular}{|c|c|c|c|c|}
\hline
region & sign of $\beta$, $\gamma$ & positive zeros  & range of $\tilde{r}$ & orbit \\
\hline\hline
I & $\beta, \gamma >0$ & 1 &
\begin{pspicture}(-3,-0.2)(2.2,0.2)
\psline[linewidth=0.5pt]{->}(-2.5,0)(1.5,0)
\psline[linewidth=0.5pt](-2.5,-0.2)(-2.5,0.2)
\psline[linewidth=1.2pt]{*-}(-1.5,0)(1.5,0)
\end{pspicture}
  & EO
\\
 & $\beta, \gamma <0$ & 0 & 
\begin{pspicture}(-3,-0.2)(2.2,0.2)
\psline[linewidth=0.5pt]{->}(-2.5,0)(1.5,0)
\psline[linewidth=0.5pt](-2.5,-0.2)(-2.5,0.2)
\psline[linewidth=1.2pt]{-}(-2.5,0)(1.5,0)
\end{pspicture}
  & TEO
\\  \hline
II & $\beta, \gamma >0$ & 2 & 
\begin{pspicture}(-3,-0.2)(2.2,0.2)
\psline[linewidth=0.5pt]{->}(-2.5,0)(1.5,0)
\psline[linewidth=0.5pt](-2.5,-0.2)(-2.5,0.2)
\psline[linewidth=1.2pt]{-*}(-2.5,0)(-1.5,0)
\psline[linewidth=1.2pt]{*-}(-0.1,0)(1.5,0)
\end{pspicture}
  & TBO, EO
\\
 & $\beta, \gamma <0$ & 1 &
\begin{pspicture}(-3,-0.2)(2.2,0.2)
\psline[linewidth=0.5pt]{->}(-2.5,0)(1.5,0)
\psline[linewidth=0.5pt](-2.5,-0.2)(-2.5,0.2)
\psline[linewidth=1.2pt]{*-}(-1.5,0)(1.5,0)
\end{pspicture}
  & EO
\\ \hline
III  & $\beta, \gamma >0$ & 3 &
\begin{pspicture}(-3,-0.2)(2.2,0.2)
\psline[linewidth=0.5pt]{->}(-2.5,0)(1.5,0)
\psline[linewidth=0.5pt](-2.5,-0.2)(-2.5,0.2)
\psline[linewidth=1.2pt]{*-*}(-1.8,0)(-1,0)
\psline[linewidth=1.2pt]{*-}(-0.1,0)(1.5,0)
\end{pspicture}
& BO, EO
\\
 & $\beta, \gamma <0$ & 0 & 
\begin{pspicture}(-3,-0.2)(2.2,0.2)
\psline[linewidth=0.5pt]{->}(-2.5,0)(1.5,0)
\psline[linewidth=0.5pt](-2.5,-0.2)(-2.5,0.2)
\psline[linewidth=1.2pt]{-}(-2.5,0)(1.5,0)
\end{pspicture}
  & TEO
\\ \hline
IV & $\beta, \gamma >0$ & 0 & 
\begin{pspicture}(-3,-0.2)(2.2,0.2)
\psline[linewidth=0.5pt]{->}(-2.5,0)(1.5,0)
\psline[linewidth=0.5pt](-2.5,-0.2)(-2.5,0.2)
\psline[linewidth=1.2pt]{-}(-2.5,0)(1.5,0)
\end{pspicture}
  & TEO
\\
 & $\beta, \gamma <0$ & 1 &
\begin{pspicture}(-3,-0.2)(2.2,0.2)
\psline[linewidth=0.5pt]{->}(-2.5,0)(1.5,0)
\psline[linewidth=0.5pt](-2.5,-0.2)(-2.5,0.2)
\psline[linewidth=1.2pt]{*-}(-1.5,0)(1.5,0)
\end{pspicture}
  & EO
\\ \hline
V & $\beta, \gamma >0$ & 4 & 
\begin{pspicture}(-3,-0.2)(2.2,0.2)
\psline[linewidth=0.5pt]{->}(-2.5,0)(1.5,0)
\psline[linewidth=0.5pt](-2.5,-0.2)(-2.5,0.2)
\psline[linewidth=1.2pt]{-*}(-2.5,0)(-1.8,0)
\psline[linewidth=1.2pt]{*-*}(-1.2,0)(-0.5,0)
\psline[linewidth=1.2pt]{*-}(0.2,0)(1.5,0)
\end{pspicture}
  & TBO, BO, EO
\\
 & $\beta, \gamma <0$ & 1 &
\begin{pspicture}(-3,-0.2)(2.2,0.2)
\psline[linewidth=0.5pt]{->}(-2.5,0)(1.5,0)
\psline[linewidth=0.5pt](-2.5,-0.2)(-2.5,0.2)
\psline[linewidth=1.2pt]{*-}(-1.5,0)(1.5,0)
\end{pspicture}
  & EO
\\ \hline\hline
\end{tabular}
\caption{Types of orbits  ($\epsilon=1$) in the cylindrical symmetric spacetime  in CG in the case of a positive cosmological constant $\Lambda>0$. The range of the orbits is represented by thick lines. The dots show the turning points of the orbits. The single vertical line indicates the singularity at $\tilde{r}=0$. An event horizon is not present for $\Lambda>0$. We do not display the cosmological horizon here, as it is not relevant for the orbits.}
\label{tab:cyl.orbits}
\end{center}
\end{table}

\clearpage

Let us now turn to the case of a \emph{negative cosmological constant} $\Lambda <0$. As before we will consider $\beta, \gamma >0$ and  $\beta, \gamma <0$. An event horizon is only present for $\beta, \gamma <0$. In the parametric  $\mathcal{L}$-$E^2$-diagram (Fig.~\ref{el2}), two different regions can be seen (to avoid confusion with the  $\Lambda >0$ case we name them region VI and VII). Table \ref{tab:cyl.orbits2} and the following list give all possible orbits for $\Lambda <0$. Comparing the General Relativity case $\gamma =0$ and the Conformal Gravity case $\gamma \neq 0$ we find that qualitatively the same orbit types occur in the GR case and the CG case if $\Lambda<0$ and $k=\alpha$.
\begin{enumerate}
	\item Region VI
	\begin{enumerate}
		\item $\beta, \gamma >0$: There are no positive zeros and $R^*(\tilde{r})<0$ for all $\tilde{r}\geq0$. Therefore geodesic motion is not possible.
		\item  $\beta, \gamma <0$:  $R^*(\tilde{r})$ has a single positive zero $r_1$ and $R^*(\tilde{r})>0$ for $\tilde{r}\in (0,r_1)$ . Here a terminating bound orbit exists.
	\end{enumerate}

	\item Region VII:
	\begin{enumerate}
		\item $\beta, \gamma >0$: In the case $\epsilon=1$, $R^*(\tilde{r})$ has a two positive zeros $r_1$,  $r_2$ and $R^*(\tilde{r})>0$ for $\tilde{r}\in [r_1,r_2)$. If $\epsilon=0$, then $R^*(\tilde{r})$ has a single positive zero $r_1$ and $R^*(\tilde{r})>0$ for $\tilde{r}\in [r_1,\infty)$. This means a bound orbit exist for particles, but lightlike geodesics move on an escape orbit.
		\item  $\beta, \gamma <0$: In the case $\epsilon=1$, $R^*(\tilde{r})$ has a single positive zero $r_1$ and $R^*(\tilde{r})>0$ for $\tilde{r}\in (0,r_1)$ . If $\epsilon=0$, then there are no positive zeros and $R^*(\tilde{r})>0$ for all $\tilde{r}\geq0$. So there are terminating bound orbits for particles and terminating escape orbits for lightlike geodesics.
	\end{enumerate}
\end{enumerate}

\clearpage

\begin{table}[h!]
\begin{center}
\begin{tabular}{|c|c|c|c|c|}
\hline
region & sign of $\beta$, $\gamma$ & positive zeros  & range of $\tilde{r}$ & orbit \\
\hline\hline
VI & $\beta, \gamma >0$ & 0 &
\begin{pspicture}(-3,-0.2)(2.2,0.2)
\psline[linewidth=0.5pt]{->}(-2.5,0)(1.5,0)
\psline[linewidth=0.5pt](-2.5,-0.2)(-2.5,0.2)
\end{pspicture}
  & no orbit
\\
 & $\beta, \gamma <0$ & 1 & 
\begin{pspicture}(-3,-0.2)(2.2,0.2)
\psline[linewidth=0.5pt]{->}(-2.5,0)(1.5,0)
\psline[linewidth=0.5pt](-2.5,-0.2)(-2.5,0.2)
\psline[linewidth=0.5pt,doubleline=true](-1.5,-0.2)(-1.5,0.2)
\psline[linewidth=1.2pt]{-*}(-2.5,0)(0,0)
\end{pspicture}
  & TBO
\\  \hline
VII & $\beta, \gamma >0$ & 2 & 
\begin{pspicture}(-3,-0.2)(2.2,0.2)
\psline[linewidth=0.5pt]{->}(-2.5,0)(1.5,0)
\psline[linewidth=0.5pt](-2.5,-0.2)(-2.5,0.2)
\psline[linewidth=1.2pt]{*-*}(-1.5,0)(0,0)
\end{pspicture}
  & BO
\\
 & $\beta, \gamma <0$ & 1 &
\begin{pspicture}(-3,-0.2)(2.2,0.2)
\psline[linewidth=0.5pt]{->}(-2.5,0)(1.5,0)
\psline[linewidth=0.5pt](-2.5,-0.2)(-2.5,0.2)
\psline[linewidth=0.5pt,doubleline=true](-1.5,-0.2)(-1.5,0.2)
\psline[linewidth=1.2pt]{-*}(-2.5,0)(0,0)
\end{pspicture}
  & TBO
\\ \hline\hline
\end{tabular}
\caption{Types of orbits ($\epsilon=1$) in the cylindrical symmetric spacetime in CG in the case of a negative cosmological constant $\Lambda<0$. The range of the orbits is represented by thick lines. The dots show the turning points of the orbits. The single vertical line indicates the singularity at $\tilde{r}=0$. The event horizon, which is present for $\beta, \gamma <0$, is marked by a double vertical line.}
\label{tab:cyl.orbits2}
\end{center}
\end{table}

\subsection{General case $k\neq \alpha$}

In contrast to GR, the CG case allows a wider range of parameters, namely four. Here we study the general case $k\neq \alpha$. We assume $\alpha^2>0$ so that the $z$ coordinate is spacelike.

First, we investigate the case $k^2<0$ where an event horizon does not exist and the singularity is naked. The parametric diagram is similar to Fig. \ref{el}, although there are less different regions. Taking $\alpha^2>0$ causes region II, IV and V to vanish. Therefore only the orbit types of region I and III are present. These orbit types cannot be found for $\gamma=0$.

Next, we consider the case $k^2>0$. An event horizon is only present for $\beta, \gamma <0$. Since $k^2$ and $\alpha$ have the same sign, the parametric diagram and the effective potential is similar to the case $k=\alpha$. Qualitatively we find the same regions and orbit types as shown in Fig. \ref{el2} and Tab. \ref{tab:cyl.orbits2}.

\subsection{Examples of the effective potential and orbit plots}
\begin{figure}[h]
    \centering
     \subfigure[]{
        \includegraphics[width=0.34\textwidth]{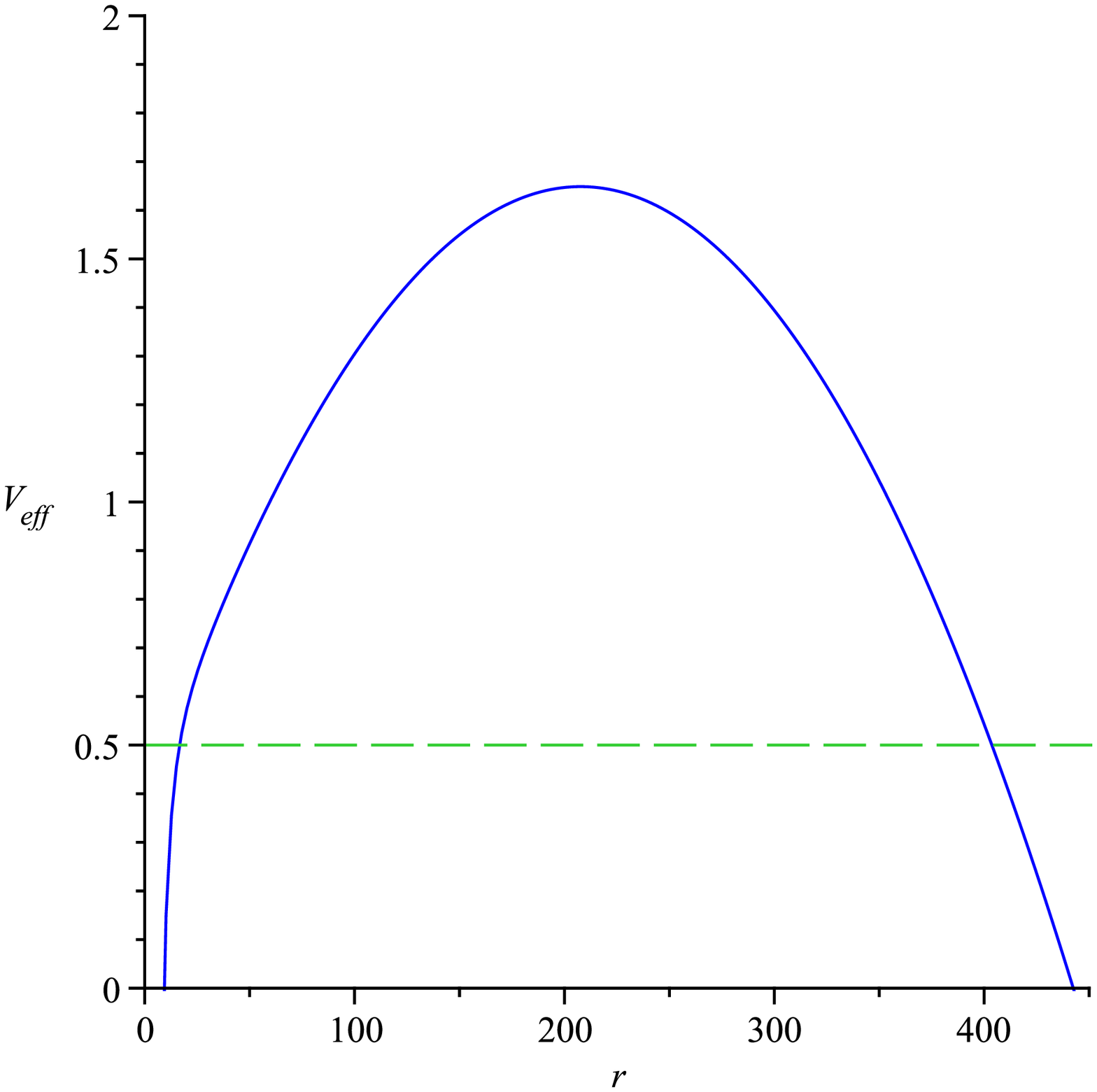}
    }
    \subfigure[]{
       \includegraphics[width=0.34\textwidth]{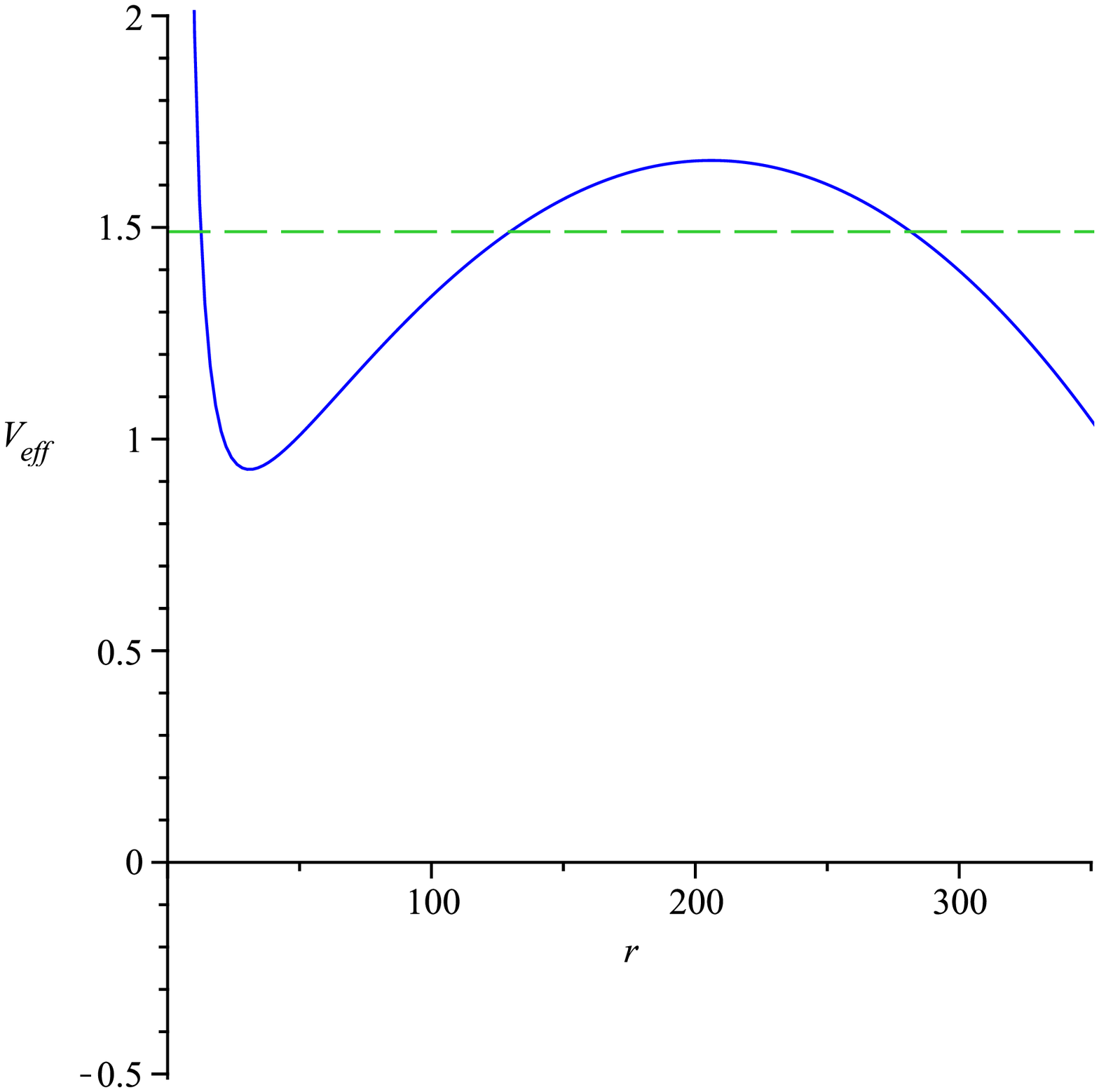}\label{v3+}
    }
    \subfigure[]{
       \includegraphics[width=0.34\textwidth]{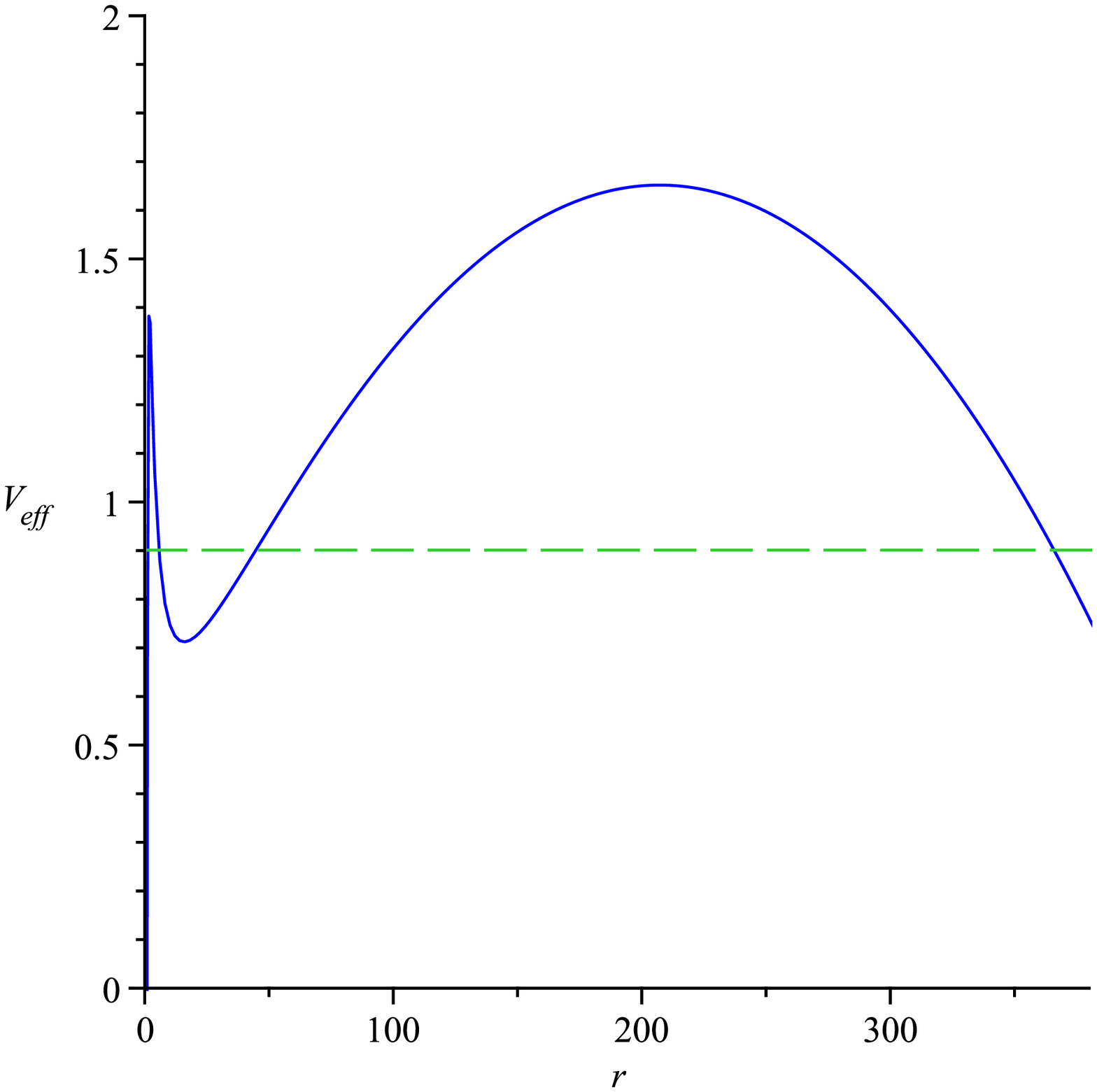}\label{v5+}
    }
     \subfigure[]{
        \includegraphics[width=0.34\textwidth]{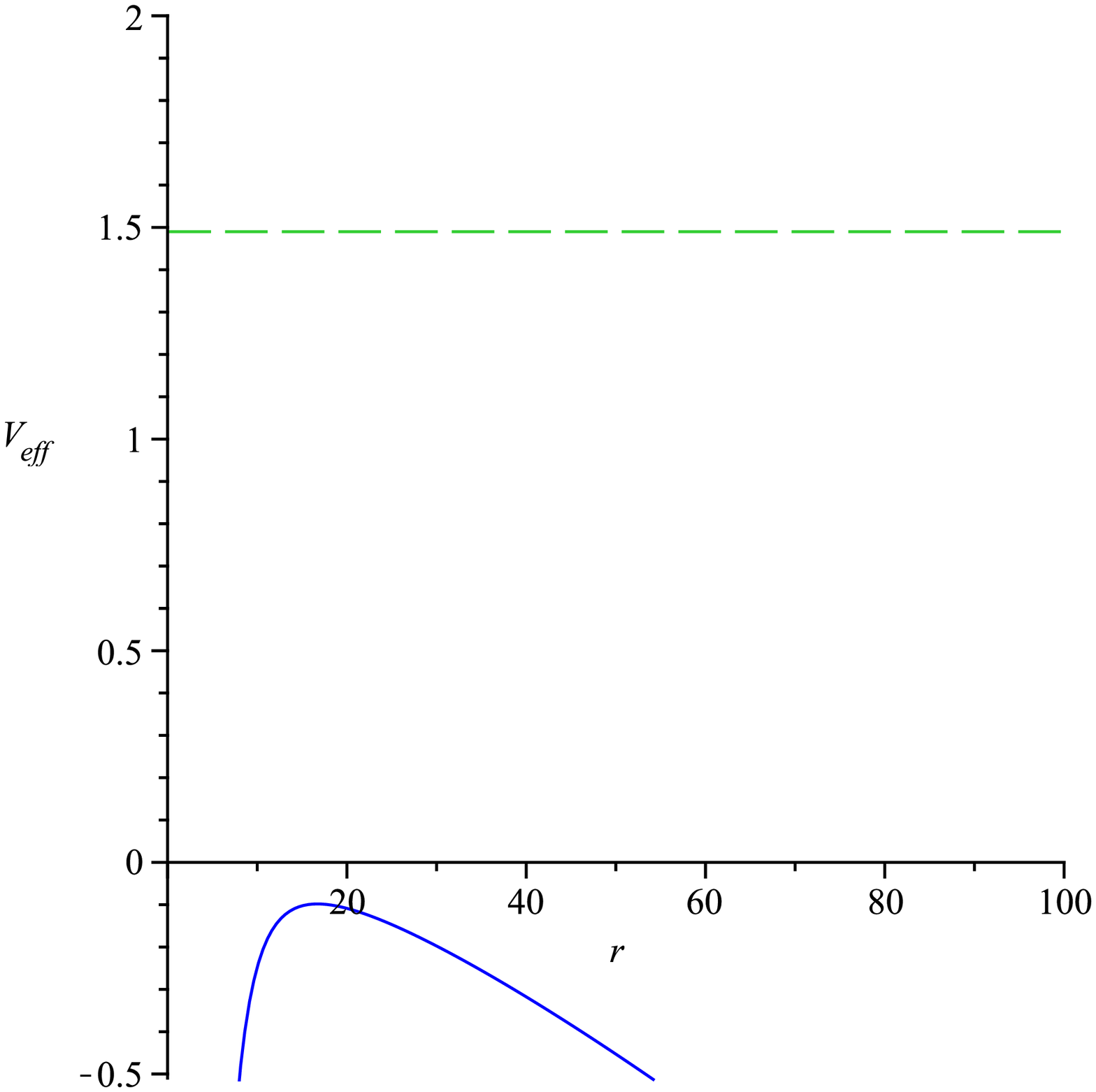}
    }
    \subfigure[]{
       \includegraphics[width=0.34\textwidth]{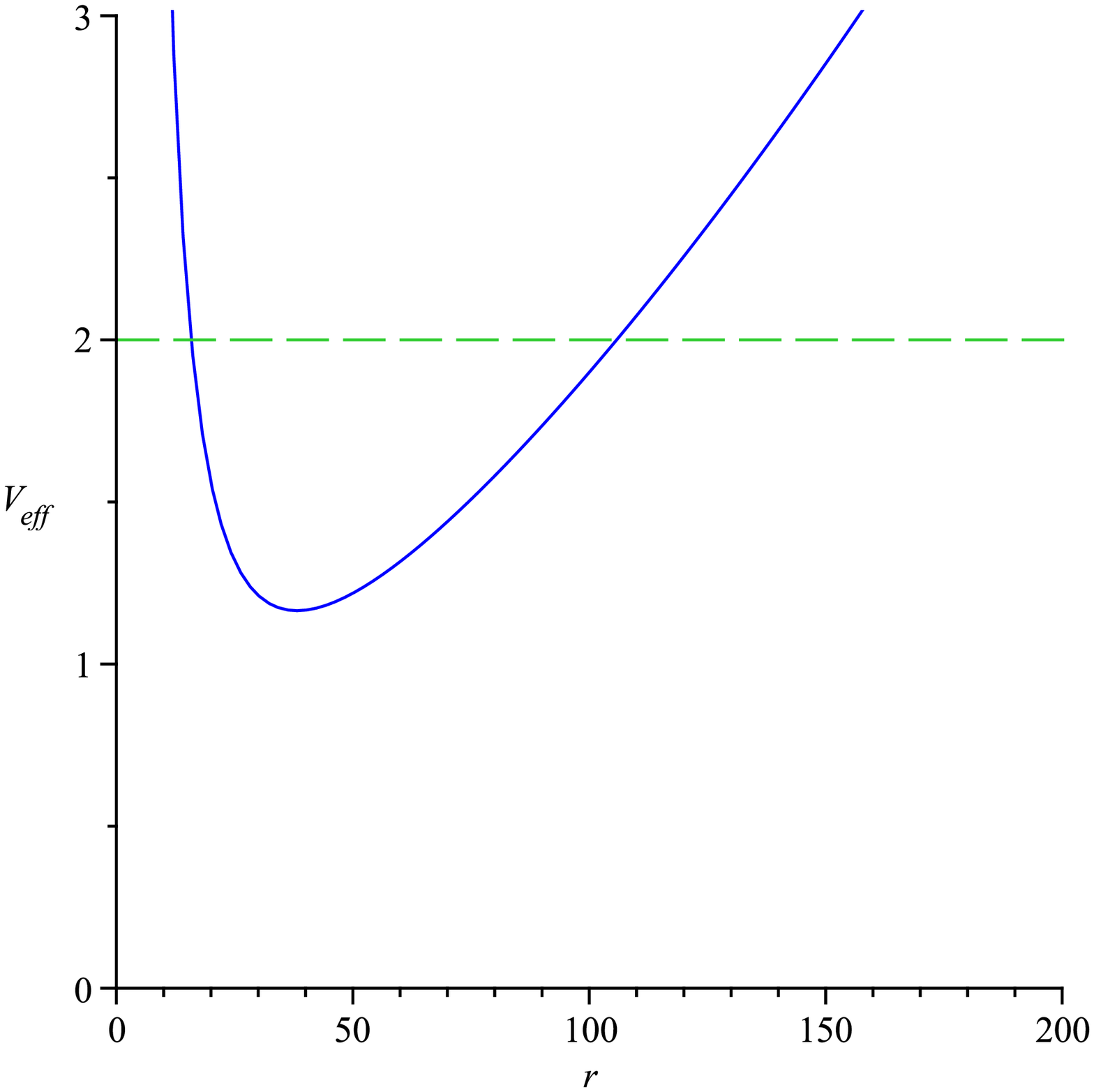}
    }
   \subfigure[]{
       \includegraphics[width=0.34\textwidth]{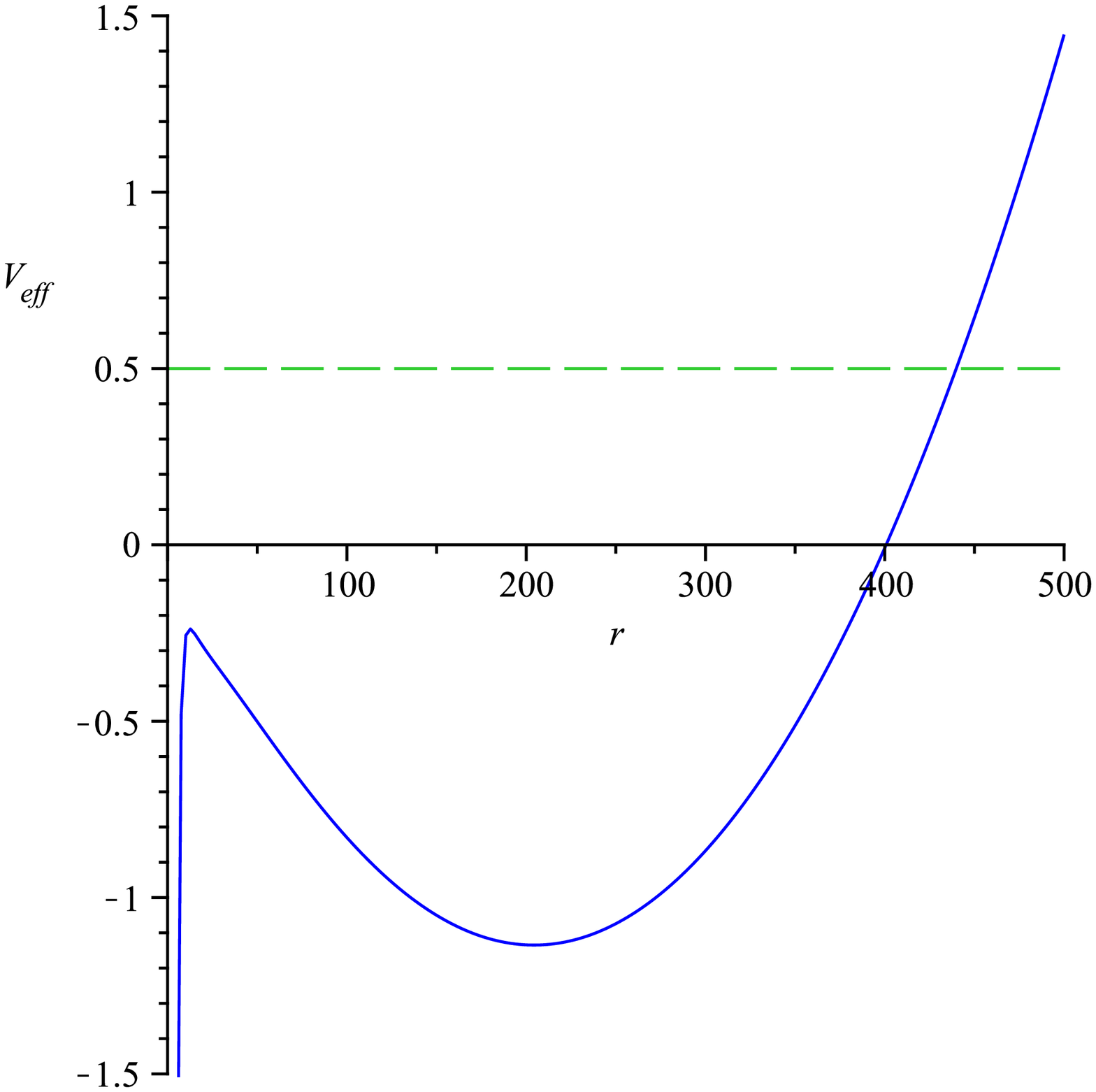}
    }
    \caption{Examples of effective potentials for geodesic motion with the parameters given in Table.~\ref{tab v}. The horizontal green dashed line represents the squared energy parameter.}
 \label{v1}
\end{figure}

\begin{figure}[h]
    \centering
     \subfigure[]{
        \includegraphics[width=0.37\textwidth]{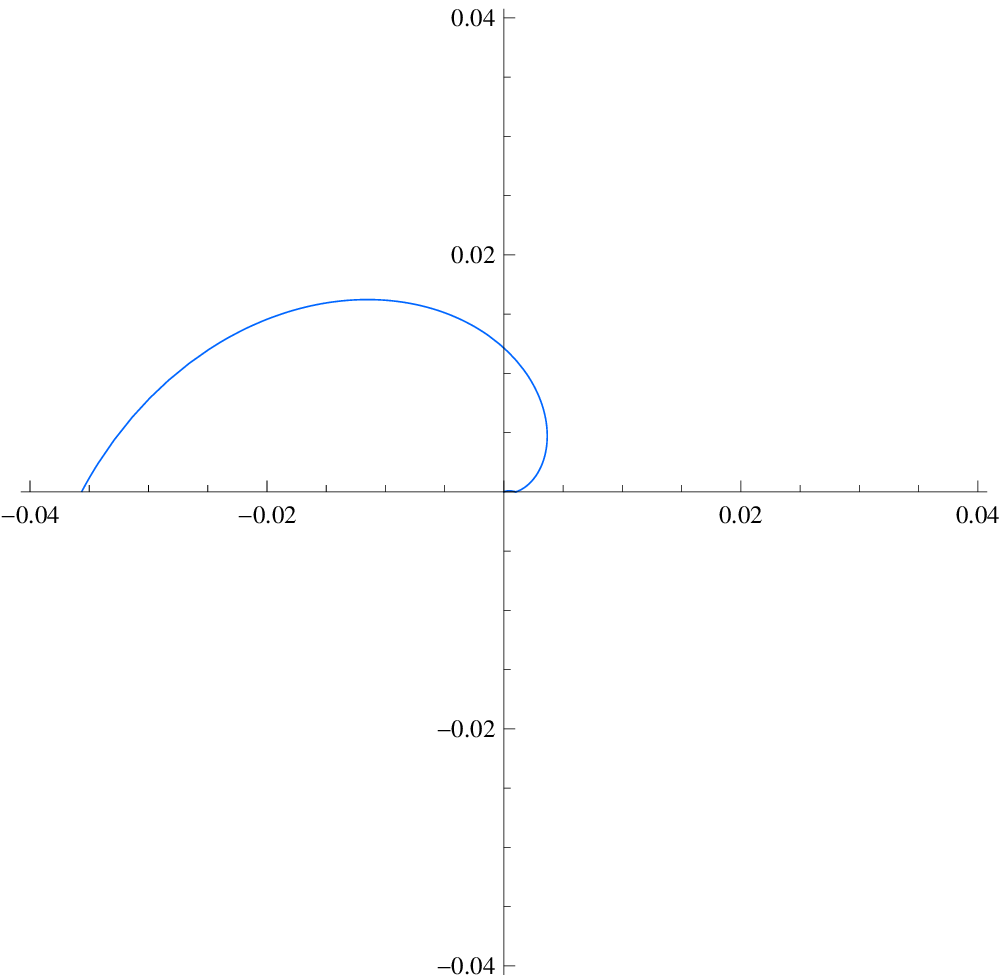}
    }
    \subfigure[]{
        \includegraphics[width=0.37\textwidth]{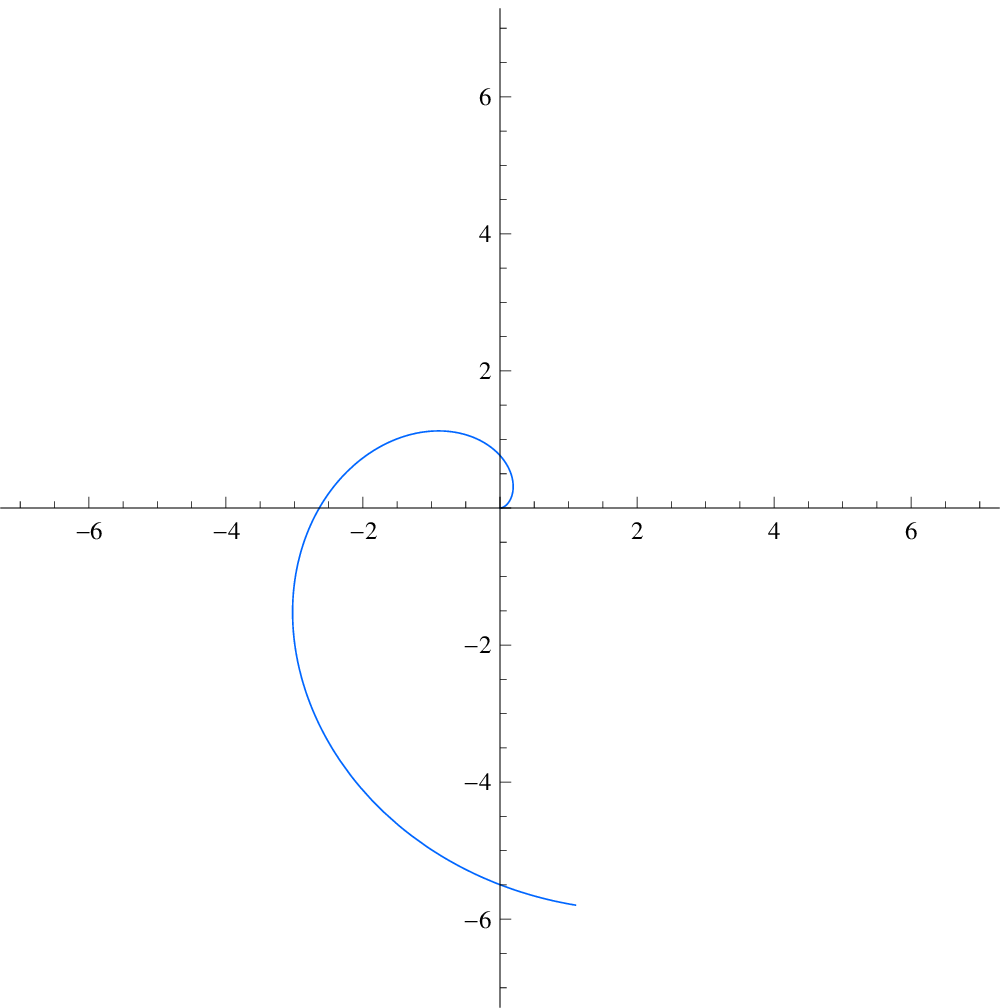}
    }
    \subfigure[]{
        \includegraphics[width=0.37\textwidth]{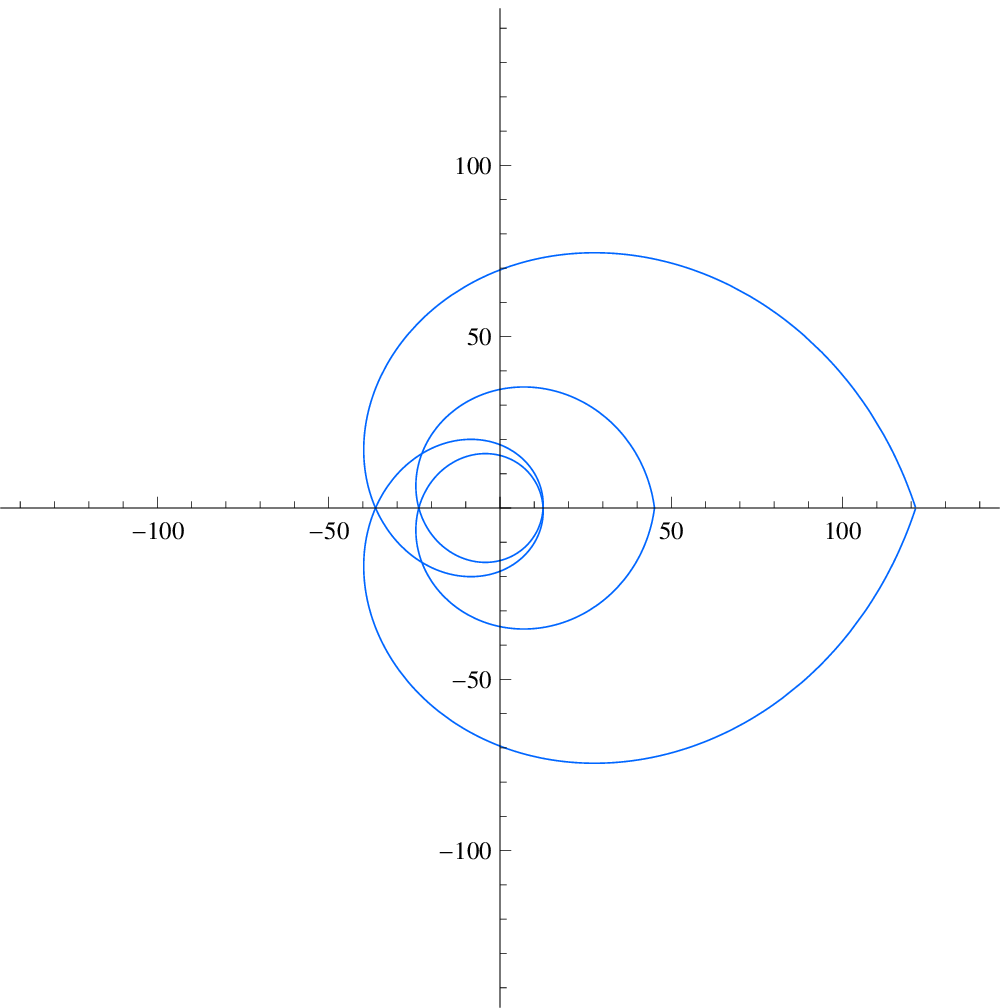}\label{o3+B}
    }
     \subfigure[]{
        \includegraphics[width=0.37\textwidth]{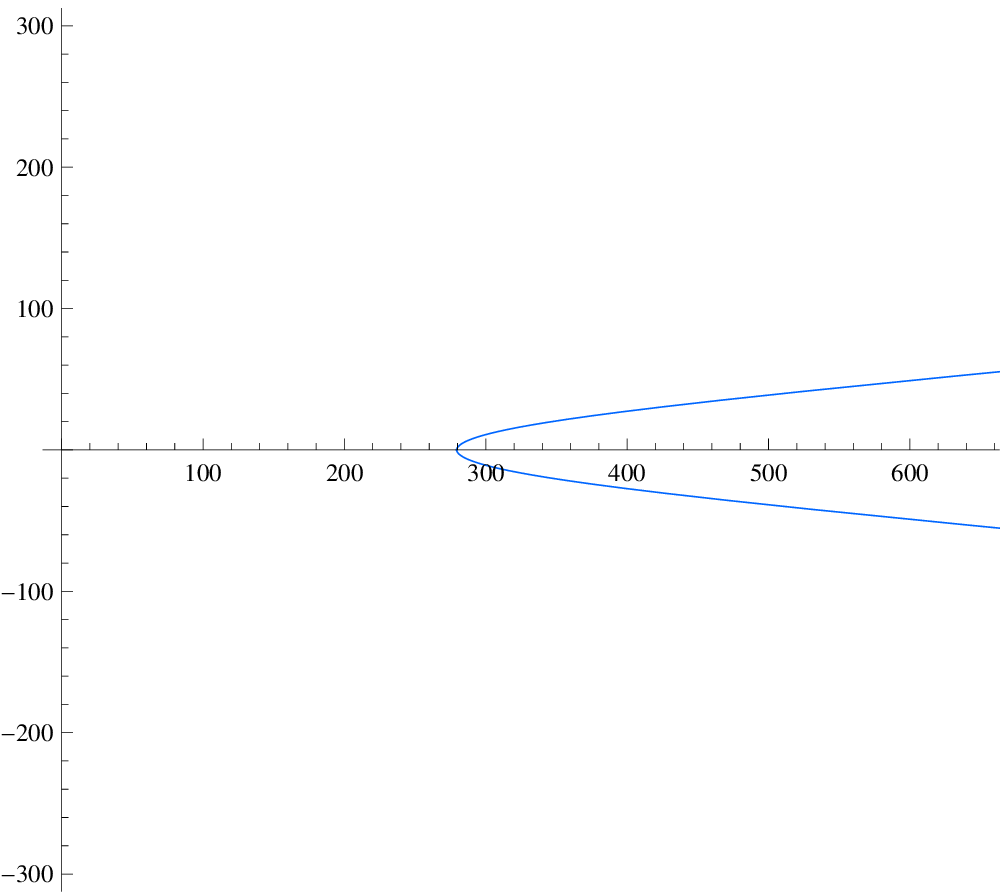}\label{o3+F}
    }
       \subfigure[]{
        \includegraphics[width=0.37\textwidth]{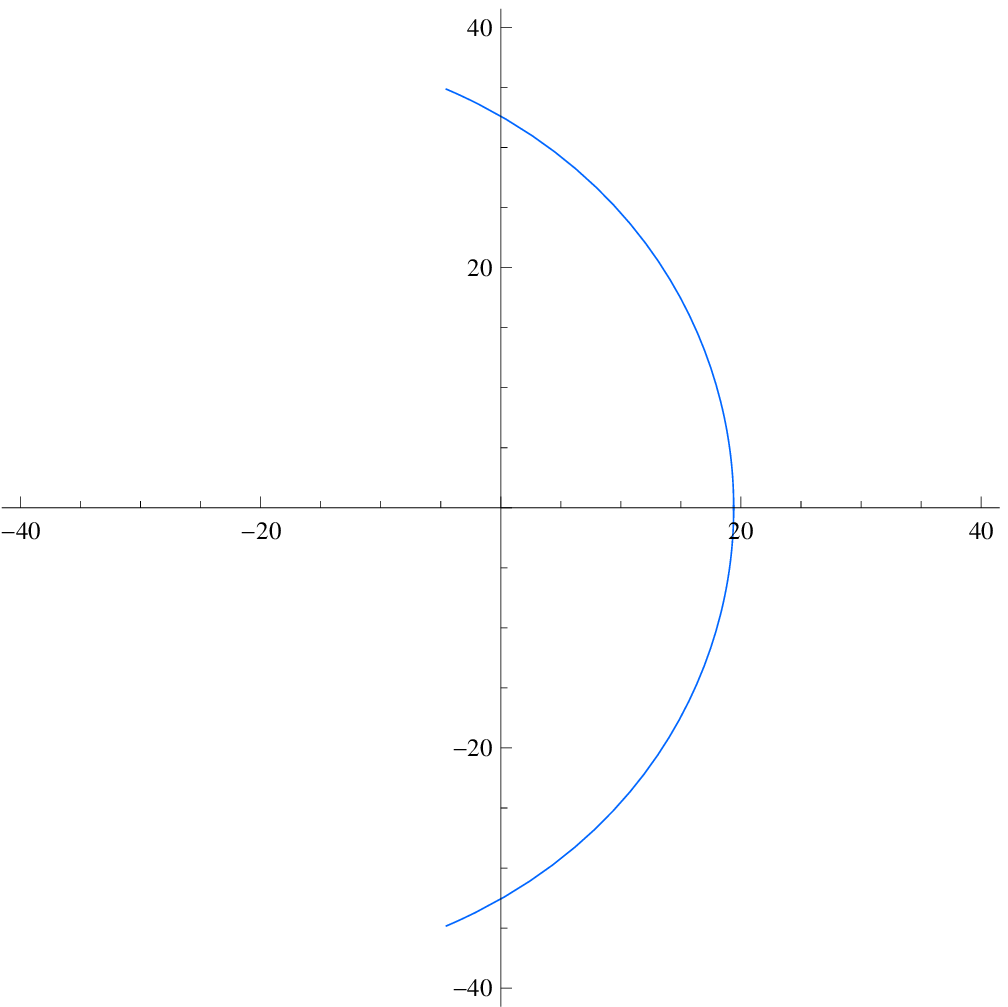}
    }
     \subfigure[]{
        \includegraphics[width=0.37\textwidth]{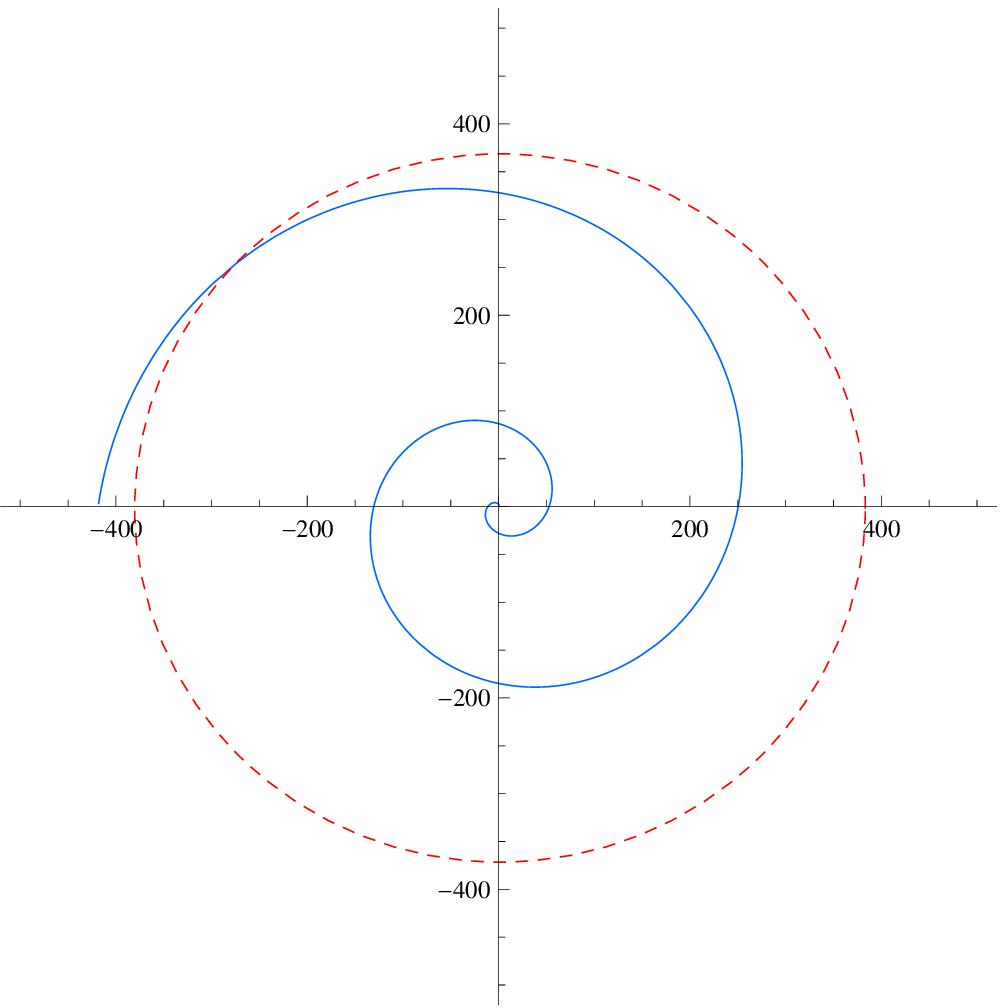}
    }
    \caption{ Example of orbit types TBO, EO, TEO, BO, EO and TBO for a, b, c, d, e and f respectively, corresponding to the Tables.~\ref{tab:cyl.orbits}--~\ref{tab v}.}
 \label{o2+}
\end{figure}

Some plots of the effective potentials for region of Fig.~\ref{el}, are shown in Fig.~\ref{v1}. Also, examples of orbit types are demonstrated in Fig.~\ref{o2+}. Note that, the effective potentials in Figs.~\ref{v3+} and \ref{v5+}, and also orbit type in Figs.~\ref{o3+B}, are not possible for GR. However, other effective potentials and orbit types are similar for GR and CG. 
\begin{table}[h]
\begin{center}
\begin{tabular}{|c|c|c|c|c|c|c|}
\hline
Fig.~\ref{v1} & $\tilde{\beta}$& $\tilde{\gamma}$   &  $\tilde{\alpha}^2=\tilde{k}^2$ & $\mathcal{L}$ & $E^2$ & region \\
\hline\hline
a & $ 3$&$ 0.05$  & $-3\cdot 10^{-5}$ & 0.004 &   0.5 & II 
\\  \hline
b & $3$&$ 0.05$  & $-3\cdot 10^{-5}$ & 0.002 & 1.5 & III
\\   \hline
c & $3$&$0.05$  &$-3\cdot 10^{-5}$&$0.00301$ & 0.9& V
\\   \hline
d & $-3$&$ -0.05$  & $-3\cdot 10^{-5}$ &0.002 & $1.5 $& III
\\   \hline
e & 1&$ 0.05$  & $3\cdot 10^{-5}$ &0.002 & 2& VII
\\   \hline
f & $-1$&$ -0.05$  & $3\cdot 10^{-5}$ &0.002 & 0.5& VI
\\ \hline\hline
\end{tabular}
\caption{Values of different parameters used for effective potential and orbit types. Also for all figures $\epsilon=1$ and $\tilde{J}=0.1$.}
\label{tab v}
\end{center}
\end{table}

\clearpage

\clearpage

\section{CONCLUSIONS}\label{c}
In this paper, we derived the equations of motion in a static cylindrically symmetric spacetime in conformal gravity. 
The geodesic equations can be solved in terms of Weierstrass elliptic functions in the case of null geodesics,
and derivatives of Kleinian sigma functions in the case of timelike geodesics. Using effective potential techniques and
parametric diagrams, we studied the possible types of orbits, which are bound orbits, escape orbits or terminating orbits. The analytic solutions of this paper can be used to calculate the exact orbits and their properties. Furthermore, observables like the periastron shift of bound orbits or the light deflection of escape orbits could be calculated. Also it would be interesting to use the analytical solutions to study the shadow of  a static cylindrically symmetric black hole in conformal gravity. Another project for future work could be the solution of the equations of motion in the charged and rotating version of this black hole spacetime.

\bibliographystyle{amsplain}

\begin{thebibliography}{999}

\bibitem{Said:2012xt} 
  J.~L.~Said, J.~Sultana and K.~Z.~Adami,
  Phys.\ Rev.\ D {\bf 85}, 104054 (2012)
  [arXiv:1201.0860 [gr-qc]].

\bibitem{Mannheim:2011ds} 
  P.~D.~Mannheim,
  Found.\ Phys.\  {\bf 42}, 388 (2012)
  [arXiv:1101.2186 [hep-th]].

\bibitem{Bender:2007wu} 
  C.~M.~Bender and P.~D.~Mannheim,
  Phys.\ Rev.\ Lett.\  {\bf 100}, 110402 (2008)
  [arXiv:0706.0207 [hep-th]].

\bibitem{Maldacena:2011mk} 
  J.~Maldacena,
  arXiv:1105.5632 [hep-th].

\bibitem{Mannheim:1988dj} 
  P.~D.~Mannheim and D.~Kazanas,
  Astrophys.\ J.\  {\bf 342}, 635 (1989).
  
\bibitem{Mannheim:1990ya} 
  P.~D.~Mannheim and D.~Kazanas,
  Phys.\ Rev.\ D {\bf 44}, 417 (1991).

\bibitem{Mannheim:2012qw} 
  P.~D.~Mannheim and J.~G.~O'Brien,
  J.\ Phys.\ Conf.\ Ser.\  {\bf 437}, 012002 (2013)
  [arXiv:1211.0188 [astro-ph.CO]].

\bibitem{Mannheim:2005bfa} 
  P.~D.~Mannheim,
  Prog.\ Part.\ Nucl.\ Phys.\  {\bf 56}, 340 (2006)
  [astro-ph/0505266].



\bibitem{Hioki:2009na} 
  K.~Hioki and K.~i.~Maeda,
  Phys.\ Rev.\ D {\bf 80}, 024042 (2009)
  [arXiv:0904.3575 [astro-ph.HE]].

\bibitem{Barack:2006pq} 
  L.~Barack and C.~Cutler,
  Phys.\ Rev.\ D {\bf 75}, 042003 (2007)
  [gr-qc/0612029].

\bibitem{Damour:1999cr} 
  T.~Damour, P.~Jaranowski and G.~Schaefer,
  Phys.\ Rev.\ D {\bf 62}, 044024 (2000)
  [gr-qc/9912092].

\bibitem{Abbott:2016blz} 
  B.~P.~Abbott {\it et al.} [LIGO Scientific and Virgo Collaborations],
  Phys.\ Rev.\ Lett.\  {\bf 116}, no. 6, 061102 (2016)
  [arXiv:1602.03837 [gr-qc]].
  
\bibitem{PerezGiz:2008yq} 
  G.~Perez-Giz and J.~Levin,
  Phys.\ Rev.\ D {\bf 79}, 124014 (2009)
  [arXiv:0811.3815 [gr-qc]].
  
  
\bibitem{Y. Hagihara}
Y. Hagihara, Japan. J. Astron. Geophys. {\bf 8}, 67 (1931).

\bibitem{S. Chandrasekhar}
S. Chandrasekhar,\textit{ The Mathematical Theory of Black
Holes} (Oxford University Press, Oxford, 1983).

\bibitem{Hackmann:2008zza} 
  E.~Hackmann and C.~L\"ammerzahl,
  Phys.\ Rev.\ Lett.\  {\bf 100}, 171101 (2008)
  [arXiv:1505.07955 [gr-qc]].
  
\bibitem{Hackmann:2008zz} 
  E.~Hackmann and C.~L\"ammerzahl,
  Phys.\ Rev.\ D {\bf 78}, 024035 (2008)
  [arXiv:1505.07973 [gr-qc]].
  
  \bibitem{Enolski:2010if}
 V.~Z.~Enolski, E.~Hackmann, V.~Kagramanova, J.~Kunz and C.~L\"ammerzahl,
  J.\ Geom.\ Phys.\  {\bf 61}, 899 (2011)
  [arXiv:1011.6459 [gr-qc]].
\bibitem{Kagramanova:2012hw} 
  V.~Kagramanova and S.~Reimers,
  Phys.\ Rev.\ D {\bf 86}, 084029 (2012);
  V.~Diemer, J.~Kunz, C.~L\"ammerzahl and S.~Reimers,
  Phys.\ Rev.\ D {\bf 89}, 124026 (2014).
\bibitem{Hackmann:2009rp}
E.~Hackmann, B.~Hartmann, C.~L\"ammerzahl and P.~Sirimachan,
  Phys.\ Rev.\ D {\bf 81}, 064016 (2010)
  [arXiv:0912.2327 [gr-qc]].
\bibitem{Hackmann:2010ir}
  E.~Hackmann, B.~Hartmann, C.~L\"ammerzahl and P.~Sirimachan,
  Phys.\ Rev.\ D {\bf 82}, 044024 (2010)
  [arXiv:1006.1761 [gr-qc]].
\bibitem{Grunau:2012ai}
 S.~Grunau, V.~Kagramanova, J.~Kunz and C.~L\"ammerzahl,
  Phys.\ Rev.\ D {\bf 86}, 104002 (2012)
  [arXiv:1208.2548 [gr-qc]].
\bibitem{Grunau:2012ri}
  S.~Grunau, V.~Kagramanova and J.~Kunz,
  Phys.\ Rev.\ D {\bf 87}, 044054 (2013)
  [arXiv:1212.0416 [gr-qc]].
\bibitem{Grunau:2013oca}
 S.~Grunau and B.~Khamesra,
  Phys.\ Rev.\ D {\bf 87}, 124019 (2013)
  [arXiv:1303.6863 [gr-qc]].

  
  \bibitem{Soroushfar:2015wqa} 
  S.~Soroushfar, R.~Saffari, J.~Kunz and C.~L\"ammerzahl,
  Phys.\ Rev.\ D {\bf 92}, 044010 (2015)  [arXiv:1504.07854 [gr-qc]].  
  
\bibitem{Soroushfar:2015dfz} 
  S.~Soroushfar, R.~Saffari and A.~Jafari,
 [arXiv:1512.08449 [gr-qc]].
  
\bibitem{Soroushfar:2016yea} 
  S.~Soroushfar, R.~Saffari and E.~Sahami,
  [arXiv:1601.03143 [gr-qc]].
  
\bibitem{Sultana:2012qp} 
J.~Sultana, D.~Kazanas and J.~L.~Said, Phys.Rev.D {\bf 86}, 084008 (2012).      

\bibitem{Varieschi:2014ata} 
  G.~U.~Varieschi,
  Gen.\ Rel.\ Grav.\  {\bf 46}, 1741 (2014)
  [arXiv:1401.6503 [gr-qc]].

\bibitem{Barabash:1999bj}
 O.V.Barabash and Y.V.Shtanov, Phys.Rev.D {\bf 60}, 064008 (1999),
  [astro-ph/9904144].
  
\bibitem{Edery:2001at} 
  A.~Edery, A.~A.~Methot and M.~B.~Paranjape,
  Gen.\ Rel.\ Grav.\  {\bf 33}, 2075 (2001)
  [astro-ph/0006173].
  
\bibitem{Wood:2001ve} 
  J.~Wood and W.~Moreau,
  gr-qc/0102056.
  
\bibitem{Brihaye:2009ef} 
  Y.~Brihaye and Y.~Verbin,
  Phys.\ Rev.\ D {\bf 80}, 124048 (2009)
  [arXiv:0907.1951 [gr-qc]].
  
\bibitem{Ohanian:2015wva} 
  H.~C.~Ohanian,
  [arXiv:1502.00020 [gr-qc]].
  
\bibitem{Lemos:1994xp} 
  J.~P.~S.~Lemos,
  Phys.\ Lett.\ B {\bf 353}, 46 (1995)
  [gr-qc/9404041].
  
\bibitem{Linet:1986sr} 
  B.~Linet,
  J.\ Math.\ Phys.\  {\bf 27}, 1817 (1986).
  
\bibitem{Santos:1993}
N.~O.~Santos,
Class. Quantum Grav. {\bf 10}, 2401 (1993).

  
\bibitem{Brihaye:2009xc} 
  Y.~Brihaye and Y.~Verbin,
  Phys.\ Rev.\ D {\bf 81}, 124022 (2010)
  [arXiv:0912.4669 [gr-qc]].

\bibitem{Verbin:2010tq} 
  Y.~Verbin and Y.~Brihaye,
  Gen.\ Rel.\ Grav.\  {\bf 43}, 2847 (2011)
  [arXiv:1008.1170 [gr-qc]].

  
  

\bibitem{M.Abramowitz}
M. Abramowitz and I. E. Stegun, \textit{Handbook of Mathematical
Functions},  (Dover Publications,New York,1968).
\bibitem{E. T. Whittaker}
E. T. Whittaker and G. N. Watson, \textit{A course of Modern
Analysis}, (Cambrige University Press, Cambrige, 1950).

\bibitem{V. M. Buchstaber} 
  V. M. Buchstaber, V. Z. Enolskii, and D. V. Leykin, Hyperelliptic
Kleinian Functions and Applications, Reviews in
Mathematics and Mathematical Physics 10 (Gordon and
Breach, New York, 1997).

\end{thebibliography}

\end{document}